\documentclass[preprint,12pt]{elsarticle}

\usepackage{amssymb}
\usepackage{amsmath}
\usepackage{xcolor}
 \usepackage{array,multirow,graphicx}
 \usepackage{setspace}
 \usepackage{hyperref}
\hypersetup{
   colorlinks,
    citecolor=black,
    filecolor=black,
    linkcolor=black,
    urlcolor=black}
    
\journal{Journal of Computational Physics}

\begin{document}

\begin{frontmatter}



\title{Fast Asymptotic-Numerical Method For Coarse Mesh Particle Simulation In Channels Of Arbitrary Cross Section}


\author[label1]{Samuel Christensen \corref{samemail}}
\author[label2]{Raymond Chu}
\author[label2] {Christopher Anderson}
\author[label1,label2]{Marcus Roper}
\cortext[samemail]{To whom correspondence should be addressed: sam.em.chris@gmail.com}

\address[label1]{Dept. of Computational Medicine, Box 951766, Life Sciences Bldg., Los Angeles, California, CA 90095-1766}
\address[label2]{Dept. of Mathematics, Box 951555, Mathematical Sciences Bldg. Los Angeles, CA 90095-1555}
\begin{abstract}
 Particles traveling through inertial microfluidic devices migrate to focusing streamlines. We present a numerical method that calculates migration velocities of particles in inertial microfluidic channels of arbitrary cross section by representing particles by singularities. Refinements to asymptotic analysis are given that improve the regularity of the singularity approximation, making finite element approximations of flow and pressure fields more effective. Sample results demonstrate that the method is computationally efficient and able to capture bifurcations in particle focusing positions due to changes in channel shape and Reynolds number.
\end{abstract}




\end{frontmatter}

\section{Introduction}
The micron-scale channels of microfluidic devices allow for manipulation of nano-liter volumes of fluid and small particles, invluding cells. At high enough fluid velocities, corresponding to Reynolds numbers between 1 and 100, interactions between particles and flow cause particles to move across streamlines, a phenomenon known as \emph{inertial focusing} \cite{di2007continuous}. Inertial focusing is finding increasingly wide use in biotechnology: differential focusing of differently-sized particles can be used to separate (large) cancer cells from (small) blood cells  \cite{che2016classification,dhar2018evaluation,chen2012microfluidic}. Meanwhile, additional hydrodynamic interactions between inertially-focused particles cause them to organize into regularly spaced trains, aiding with their encapsulation or analysis\cite{lee2010dynamic,hood2018pairwise}.
Numerical simulation of inertial microfluidic devices requires solving for the 3D motion of suspended in a fluid flow occurring in a 3D geometry that is changing due to the movement of the particles.  Moreover, the phenomenon of interest, particle migration, and in particular particle focusing occur at non-negligible Reynolds number, so cannot be modeled by numerical methods that are designed for small Reynolds number particle-flows such as Stokesian dynamics \cite{sierou2001accelerated} or boundary integral methods \cite{liron1992motion}. 

The time evolving geometry of migrating particles, and need to include nonlinear terms, favors numerical methods such as immersed boundary \cite{kazerooni2017inertial,nakagawa2015inertial} or immersed interface \cite{xu2020octree} that embed moving boundaries within a fixed computational grid on which the Navier-Stokes equations are solved. Relative easy parallelization has boosted the popularity of Lattice-Boltzmann methods (LBM) for approximating the Navier-Stokes equations \cite{bazaz2020computational}. Moving boundary LBM-based simulations have been used to study the dependences of focusing upon channel aspect ratio \cite{sun2016three}, particle size and Reynolds number \cite{chun2006inertial}, as well as to incorporate additional physics due to deformable particles \cite{schaaf2019flowing}, or interactions between multiple particles \cite{humphry2010axial,kahkeshani2016preferred}. Yet, although existing numerical simulations have illuminated the physics of inertial focusing, the high computational cost of nonlinear 3D simulations has meant that predictive simulations are not currently used to design or optimize inertial microfluidic devices.

Here we present a numerical method based on the singular asymptotic expansions associated with flows where the ratio of particle size to domain size is small \cite{schonberg1989inertial}. When a suspended particle is small compared to the size of the domain, the fluid-filled domain may be divided into near- and far-field regions. Flow fields can be computed analytically within the near-field region, while in the far-field region the Navier-Stokes Equations are linearized and then solved numerically. By extending the asymptotic expansion of the near field solution to include higher order terms, we increase the regularity of the numerical far-field solution from discontinuous to continuous, which allows for second order convergence of a polynomial based finite element solution of the far field equations. Thus, the far-field solution is solved rapidly and accurately using polynomial finite elements. Although the method is generalizable to multiple particles and to fully three dimensional domains, we limit our attention here to predicting all particle focusing positions and focusing trajectories within channels of arbitrary cross-sections, a problem for which existing experimental and simulation data allows for testing of predictions. An implementation of the numerical method in Matlab may be downloaded from \href{https://samuelechristensen.github.io/InFocus/}{GitHub}, and can be run on computers with as little as 8GB of memory.

An existing method that also combines asymptotic and numerical techniques is the PHYSALIS resolved-particle simulation code \cite{prosperetti2001physalis}. The PHYSALIS simulation is based on a model that assumes in the rest frame of a rigid particle the velocity of the fluid close to the particle surface is close to zero. Leading order fluid acceleration terms can be absorbed into pressure, so in the near particle region, the fluid flow equations obey approximately the Stokes equations. The PHYSALIS code uses exact Stokes solutions to model fluid flow near  particles and numerically couples the near field solutions to fully nonlinear fluid flow away from the particles. It can solve, in principle, for any particle shape where exact Stokes solutions are available, including for multiple spherical particles or 2D elliptical particles \cite{prosperetti2001physalis,sierakowski2016resolved}. However the code still requires a fully resolved 3D Navier-Stokes simulation to solve for the flow away from particle neighborhoods. Rigorous analysis of the effects of where and how matching between Stokes and Navier-Stokes solutions remains an ongoing challenge. It is highly likely that the existing PHYSALIS code base could be used to model particle focusing in rectangular channels. The principal gains from the method described in this paper are that linear approximation in the outer region allows the outer field to be calculated using a single fast linear solve, and our use of finite element methods allows a diverse range of channel shapes to be modeled. 



\label{methods}
\section{Derivation of Approximate Equations Of Motion}
\label{section:EquationsOfMotion}
\begin{figure}
    \begin{center}
    \includegraphics[width=0.9\textwidth]{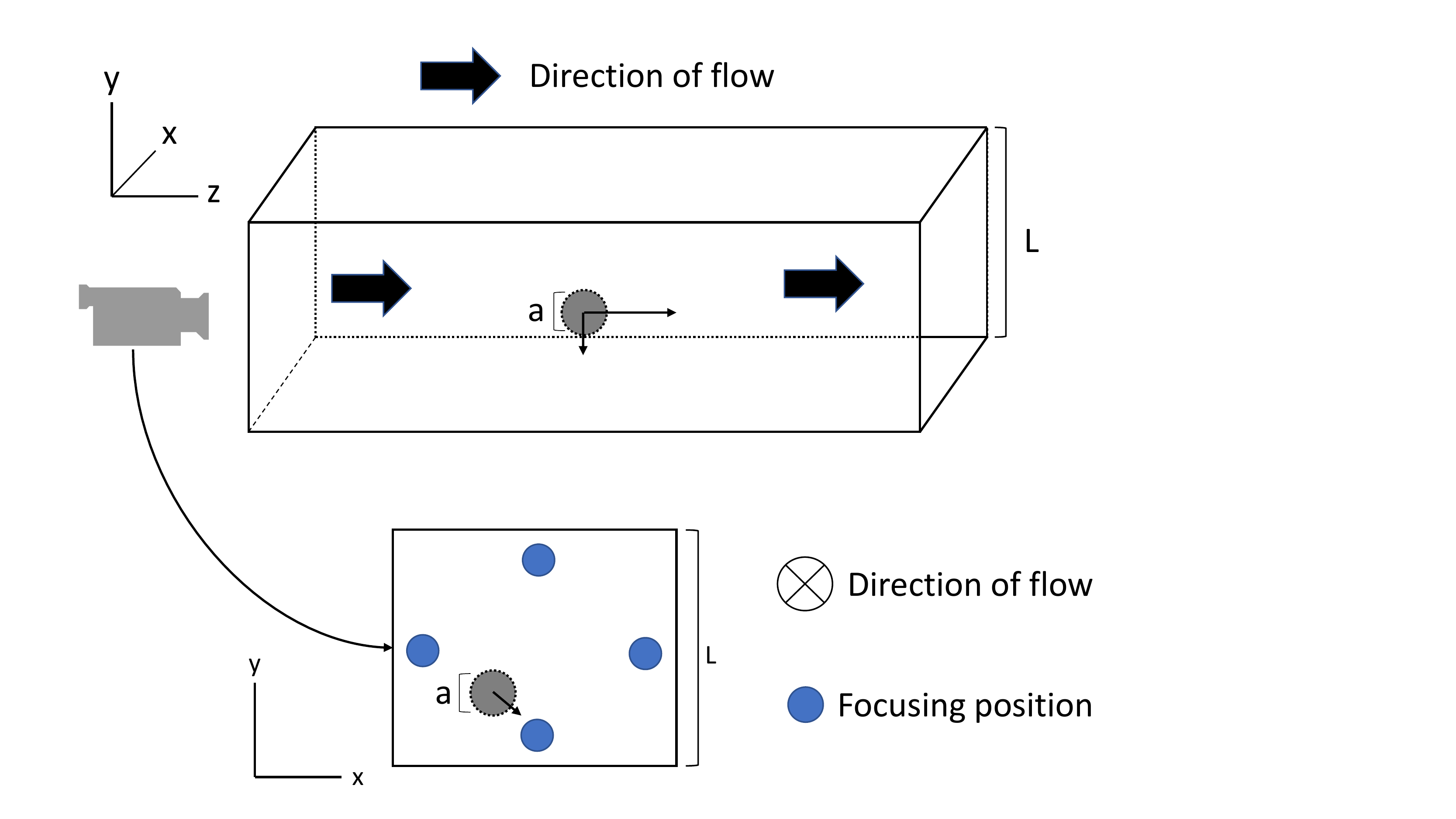}
    \caption{Schematic showing sphere of radius $a$, freely flowing in a straight channel of uniform cross-section shape. Propelled by inertial migration forces, the particle converges to one of several focusing streamlines, shown in the cross-section view.}
    \label{fig:schematic}
    \end{center}
\end{figure}

The problem considered in this paper is the prediction of particle trajectories in a straight microfluidic channel, with arbitrary cross-section. Due to inertial focusing forces, spherical particles introduced within the channel migrate across streamlines, arriving ultimately at one of a small number of inertial focusing locations (Fig. \ref{fig:schematic}). We present a numerical method for predicting the focusing trajectory of particles, as well as the number of possible final focusing position and their basins of attraction.

Two Reynolds numbers can be defined for the flow: the channel Reynolds number $Re_c = \frac{\rho U_{\rm max}L}{\mu}$, constructed from the fluid viscosity, $\mu$, density $\rho$, maximum velocity $U_{\rm max}$ and a channel length-scale (e.g. diameter), $L$. In typical microfluidic devices $Re_c$ values are in the tens or hundreds. The particle Reynolds number $Re_p = \alpha^2 Re_c$, describes the velocity disturbances created by the particle, and involves an additional factor $\alpha \equiv \frac aL$ which is often much less than 1, so that $Re_p$ is $O(1)$ or smaller.  

We decompose fluid velocities within the microfluidic channel into two components: the flow without the particle (also called the background flow), $\bar{\mathbf{u}}$, and the disturbance velocity created by the particle: $\mathbf{u'}$. The total fluid velocity is $\mathbf{u}=\bar{\mathbf{u}}+\mathbf{u'}$. Furthermore, we introduce dimensionless variables $\mathbf{u}^*=\frac{\mathbf{u}'L}{aU_{max}}, \ \mathbf{x}^*=\frac{\mathbf{x}}{a}, \ 
t^*=\frac{t U_{max}}{L}, \ p^*=\frac{Lp}{\mu U_{max}}$, and we change the frame of reference so that $\mathbf{x}_p$, the center of our moving particle is now the origin.  The equations for the dimensionless disturbance velocity are:
\begin{align} \label{eq:NSfinal}
\Delta \mathbf{u}^* - \nabla p^* &= Re_p\left(\frac{\partial \mathbf{u}^*}{\partial t^*}+\left(\mathbf{U_p^*}(t^*)+\bar{\mathbf{u}}(\mathbf{x}_p)+\mathbf{u}^*\right)\cdot \nabla \mathbf{u}^*+\mathbf{u}^*\cdot\nabla \bar{\mathbf{u}}'+ \bar{\mathbf{u}}'\cdot \nabla \mathbf{u}^*\right) \nonumber\\
\nabla \cdot \mathbf{u}^* &= 0 \nonumber \\
\mathbf{u}^* &= \mathbf{0}  \qquad \text{    on the channel walls} \nonumber \\
\mathbf{u}^* &= \mathbf{U}_p^*(t^*) + \boldsymbol{\Omega}_p(t^*)\times \mathbf{x}  \qquad\text{on }|\mathbf{x}|=1
\end{align}
where $\bar{\mathbf{u}}'(\mathbf{x}) = \bar{\mathbf{u}}(\mathbf{x})-\bar{\mathbf{u}}(\mathbf{x}_p)$ is the background flow in frame of reference with the particle. The particle velocity and rotation, $\mathbf{U}^*_p$ and $\boldsymbol{\Omega}^*_p$ evolve over time to satisfy the force and torque free condition. The $x$ and $y$ components of $\mathbf{U}^*_p$ constitute the particle's migration velocity. Henceforth we drop the asterisks distinguishing dimensionless from dimensional quantities.

 The asymptotic approximations incorporated into our numerical method are derived from the system of equations \ref{eq:NSfinal} in the limit $\alpha\to 0$ and $Re_p\to 0$. In these limits, flow field separates into an {\bf inner region}, where the effects of the particle dominate but where inertial terms can be neglected at leading order, and we can expand the velocity field: $\mathbf{u} \sim \alpha Re_p^0 \mathbf{u}_1^{(0)} + \alpha Re_p \mathbf{u}_1^{(1)}+\ldots$  and an {\bf outer region} where the inertia of the outer region interacts with the disturbance coming from the inner region to create a disturbance velocity: $\mathbf{u} \sim \alpha Re_p \mathbf{U}_1^{(1)} + \ldots$ \cite{van1975perturbation, proudman1957expansions, kaplun1957asymptotic, schonberg1989inertial}.  We similarly expand the particle's translational velocity as $\mathbf{U}_p=\alpha Re_p^0 \mathbf{U}_{p,0}^{(0)} + \alpha Re_p^1 \mathbf{U}_{p,1}^{(1)}\ldots$ and the rotation speed as $\boldsymbol{\Omega}_p = \alpha Re_p^0\boldsymbol{\Omega}_{p,1}^{(0)}+\alpha Re_p^1\boldsymbol{\Omega}_{p,1}^{(1)}+\ldots$.

\subsection{Inner Region Expansion of the Equations of Motion}
In the inner region we scale lengths by $a$, the particle radius, and extract the order $O(\alpha Re_p^0)$ terms in Eqn. \ref{eq:NSfinal} to describe the disturbance velocity near the particle \cite{schonberg1989inertial}:
\begin{align}
    \Delta \mathbf{u}_1^{(0)} - \nabla p_1^{(0)} &= \boldsymbol{0} \nonumber \\
    \mathbf{u}&=  -\boldsymbol{\gamma}\cdot\mathbf{x}\mathbf{e}_3+\boldsymbol{\Omega}_{p,1}^{(0)} \times\mathbf{x}\ \ \ \text{on } |\mathbf{x}|=1  \\
    \mathbf{u}&\rightarrow \boldsymbol{0} \ \ \hbox{as} \ |\mathbf{x}| \rightarrow \infty \nonumber.
\end{align}
Where $\boldsymbol \gamma$ describes the local velocity gradient, also known as the shear rate, of the background flow at the location of the particle. The rotation component is there to satisfy the torque free condition.  This equation can be solved analytically to obtain $\mathbf{u}_1^{(0)}\sim \mathbf{u}_{\rm str}+O\left(1/r^4\right)$ as $r\equiv |\mathbf{x}|\to \infty$ \cite{kim2013microhydrodynamics}; the \textit{stresslet} flow field, defined by
 \begin{align}
 \label{eq:stresslet}
 \mathbf{u}_{\rm str} = -\frac{1}{r^2}\bigg(\gamma_x \frac{5xz \mathbf{x}}{2r^3} + \gamma_y \frac{5yz \mathbf{x}}{2r^3}\bigg)~,
 \end{align}
 represents the slowest decaying component of the disturbance velocity far away from the particle. Importantly, $\mathbf{u}_1^{(0)}$ depends only on the shear rate $\boldsymbol{\gamma}$ and there is no migration velocity generated at this order \cite{cox1968lateral}.

   \subsection{Outer Region Expansion of the Equations of Motion}
 \label{sec:FarField}
The presence of the particle disturbs the background flow locally and an investigation of how this disturbance interacts with the background flow throughout the channel leads to the construction of equations for fluid motion in the outer region. Far from the particle, the rapidly decaying viscous stress term ($\Delta \mathbf{u_{\rm str}}=O\left(\frac{1}{r^4}\right)$) will decrease until it reaches equal magnitude with the more slowly decaying inertial term ($Re_p\left( \bar{\mathbf{u}}'\cdot \nabla \mathbf{u}_{\rm str}\right)=O\left(\frac{Re_p}{r^2}\right)$), at a distance $r=O(Re_p^{-1/2})$. Accordingly we define coordinates $\mathbf{X} = Re_p^{1/2}\mathbf{x} $. Generalizing the result of Sch\"onberg and Hinch\cite{schonberg1989inertial} for the $O(\alpha Re_p)$ leading order disturbance velocity to 2D channel shapes, we obtain:
 \begin{align} \label{eq:leadingOuter}
     \Delta\mathbf{U}_1^{(1)} -\nabla P_1^{(1)} - \mathbf{U}_1^{(1)} \cdot \nabla \mathbf{\bar{U}}' - \mathbf{\bar{U}}'\cdot \nabla \mathbf{U}_1^{(1)}  &= -\begin{bmatrix}
     \gamma_x \frac{\partial}{\partial Z} \\
     \gamma_y \frac{\partial}{\partial Z} \\
     \gamma_x \frac{\partial}{\partial X} +\gamma_y \frac{\partial}{\partial Y} 
     \end{bmatrix} \delta(\mathbf{X}) \\
     \nabla \cdot \mathbf{U}_1^{(1)} &= \mathbf{0} \nonumber \\
     \mathbf{U}_1^{(1)} &= \mathbf{0}  \quad \hbox{on channel walls}  \nonumber
 \end{align}
Matching of the outer region flows to the slowest decaying component of the inner region solution requires that $\mathbf{U}_1^{(1)}\sim \mathbf{u}_{\rm str}$ as $\mathbf{X}\to \mathbf{0}$. In Eqn. \ref{eq:leadingOuter}, this matching is accomplished by adding derivatives of the Dirac delta function, $\delta(\mathbf{X})$, as forcing terms \cite{schonberg1989inertial}. Once the stresslet has been subtracted, the limiting value of $\mathbf{U}_1^{(1)}$ as $\mathbf{X}\rightarrow 0$ can be used to compute the particle migration velocity at $O(Re_p)$. This PDE is much easier to solve numerically than the original system (Eqn. \ref{eq:NSfinal}) because it is linear and because there is no need to explicitly representing the moving particle boundary.
 
\subsection{Reformulating the problem for a continuous solution} 
\label{sec:discontinuityReformulation}
 \begin{figure}
     \centering
     \includegraphics[width=0.8\textwidth]{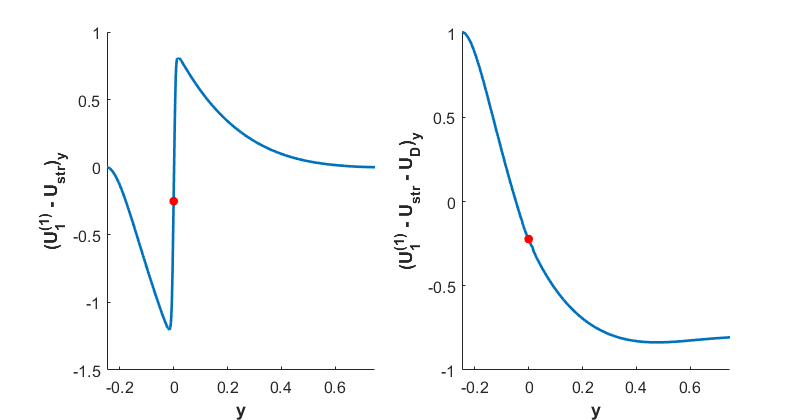}
     \caption{The velocity $\mathbf{U}_1^{(1)}-\mathbf{U}_{\rm str}$ is discontinuous, even once the stresslet contribution is subtracted. Left panel: the $Y-$component of the partially regularized velocity, $(\mathbf{U}_1^{(1)}-\mathbf{U}_{\rm str})_Y$ along the line segment $X=Z=0$ is discontinuous at $Y=0$. Right panel: We accelerate particle focusing simulations by solving for the continuous velocity field $\mathbf{V}=\mathbf{U}_1^{(1)}-\mathbf{U}_{\rm str}-\mathbf{U}_D$ using Eqn. \ref{eq:CTS}. Velocity fields agree at $Y=0$ (red dot), to give the particle migration velocity. Channel geometry is 4$\times$1$\times$10 rectangular channel with $Re_c=1$.}
     \label{fig:Discontinuity}
 \end{figure}
 
The solution of Eqn. \ref{eq:leadingOuter} is singular at $\mathbf{X}=\mathbf{0}$. The singularity can be removed by subtracting off the stresslet velocity and pressure fields (Eqn. \ref{eq:stresslet}) from $\mathbf{U}_1^{(1)}$. However, the resulting non-singular solution remains discontinuous, and the polynomial finite element basis is a computationally inefficient representation for the solution's remaining three dimensional discontinuity, limiting the method to first order convergence. By removing the discontinuity we can construct a finite element solver that achieves second order convergence. We find the discontinuity analytically by examining $\mathbf{u}_1^{(1)}$, the inner solution at $O(\alpha Re_p)$, which satisfies:
  \begin{align}
     \Delta \mathbf{u}_1^{(1)} -\nabla p_1^{(1)}&= (\boldsymbol\gamma \mathbf{\cdot \mathbf{x}\mathbf{e}_3}) \cdot \nabla \mathbf{u}_1^{(0)} +  \mathbf{u}_1^{(0)}\cdot \nabla  (\boldsymbol\gamma \mathbf{\cdot \mathbf{x}\mathbf{e}_3}) \label{eq:u11inner}
     \\\nabla \cdot \mathbf{u}_1^{(1)} &= 0~,
 \end{align}
 with appropriate velocity boundary conditions to maintain a force and torque free particle. Howecwe, these boundary conditions do not affect the non-decaying component of the velocity solution.We are only interested in the non-decaying component of this solution which is not affected by the boundary condition. There is no migration or rotation generated from the inner region due to this solution \cite{hood_lee_roper_2015, happel2012low}, but components of $\mathbf{u}_1^{(1)}$ do not decay as  $r\to \infty$ and therefore affect the leading order solution in the outer region. We solved for the nondecaying components of Eqn. \ref{eq:u11inner} analytically using spherical harmonics and the projection method (see \ref{sec:discontinuitySolve}).

In order to take advantage of \textit{a priori} knowledge of the discontinuity, we decompose $\mathbf{U}_1^{(1)}$ and $P_1^{(1)}$ into regular and irregular (singular $+$ discontinuous) components: $\mathbf{U}_1^{(1)} = \mathbf{V} + \mathbf{U}_{\rm str} + \mathbf{U}_D$ and $P_1^{(1)} = Q + P_{\rm str} + P_D$, where ($\mathbf{U}_{\rm str}$, $P_{rm str}$) is the stresslet solution from Eqn \ref{eq:stresslet} and ($\mathbf{U}_D$, $P_D$) are the discontinuous velocity fields given in Eqn. \ref{eq:discontinuitySolution}, rewritten in terms of the outer variable $\boldsymbol{X}$. Rewritten in terms of $(\mathbf{V},Q)$, Eqn. \ref{eq:leadingOuter} becomes:
\begingroup\makeatletter\def\f@size{10}\check@mathfonts
  \begin{align} \label{eq:CTS}
     \Delta \mathbf{V} -\nabla Q - \mathbf{V} \cdot \nabla \mathbf{\bar{U}} + \mathbf{\bar{U}}\cdot \nabla \mathbf{V} &=   \mathbf{U}_{\rm str} \cdot \nabla (\mathbf{\bar{U}}{-}\boldsymbol{\gamma} \cdot \mathbf{x} \mathbf{e}_3) + \mathbf{U}_{\rm str} \cdot \nabla (\mathbf{\bar{U}}{-}\boldsymbol{\gamma} \cdot \mathbf{x} \mathbf{e}_3)\nonumber  \\
     &\ \ + \mathbf{U}_D\cdot \nabla \mathbf{\bar{U}} + \mathbf{\bar{U}}\cdot \nabla \mathbf{U}_D \nonumber \\
     \nabla \cdot \mathbf{V} &= \ \mathbf{0} \\
     \mathbf{V} &= -\mathbf{U}_{\rm str}-\mathbf{U}_D \ \ \text{on walls of the channel} \nonumber
 \end{align}
 \endgroup
This reformulation increases numerical accuracy not just because the unknown velocity and pressure fields are made continuous at $\mathbf{X}=\mathbf{0}$, but because the order of the singularity in the right hand side of the first subequation in Eqn. \ref{eq:CTS} is reduced from $O(R^{-2} )$ to $O(R^{-1})$.

\subsection{Alternative Regularization Methods}
\label{sec:alternativeFormulations}
Singular forcing terms are commonly encountered in computational models of fluid-body interactions, and there are existing methods that can be used to regularize a system of equations such as Eqn. \ref{eq:leadingOuter}. We compare the regularization from Section \ref{sec:discontinuityReformulation}, with 1. a direct solve of Eqn. \ref{eq:leadingOuter} with a regularized Dirac delta function and 2. velocity and pressure decomposition, in which a regularized form of the stresslet is subtracted from $\mathbf{U}_1^{(1)}$.

To regularize the Dirac delta forcing term in Eqn. \ref{eq:leadingOuter} directly, we replace $\delta(\mathbf{X})$ by normalized Gaussian of $\epsilon-$variance, $\delta^\epsilon(\mathbf{x}) = \frac{1}{(2\pi\epsilon)^{\frac{3}{2}}}\exp \left(-\frac{||{\mathbf{x}}||^2}{2\epsilon^2}\right)$:
\begin{align} \label{eq:singular}
     \Delta\mathbf{U_1^{(1)}} -\nabla P_1^{(1)} - \mathbf{U}_1^{(1)} \cdot \nabla \mathbf{\bar{U}}' - \mathbf{\bar{U}}'\cdot \nabla \mathbf{U}_1^{(1)}  &= -\frac{10 \pi}{3}\begin{bmatrix}
     \gamma_x \frac{\partial}{\partial z} \\
     \gamma_y \frac{\partial}{\partial z} \\
     \gamma_x \frac{\partial }{\partial x} +\gamma_y \frac{\partial}{\partial y} 
     \end{bmatrix} \delta^\epsilon(\mathbf{x})  \\
     \nabla \cdot \mathbf{U}_1^{(1)} &= \mathbf{0} \nonumber
\end{align}

A second regularization method is modeled on the regularized Stokeslet method for blunting point force singularities in zero-$Re$ flows \cite{cortez2001method}. Specifically, we isolate the singular part of the velocity field; $\mathbf{U}_0^{(1)} = \mathbf{U}_{\rm str} + \mathbf{V}_{ns}$ where $\mathbf{U}_{\rm str}$ is the stresslet written in terms of the outer coordinates. We present a regularized version of $\mathbf{U}_{\rm str}$ that is spread over nearby grid points, $\mathbf{U}_{\rm str}^\epsilon$, by solving Stokes's equations with delta function forcing by a blob with length scale $\epsilon$ \cite{cortez2001method}:
\begin{align} \label{regStresslet}
    \Delta \mathbf{U}^\epsilon_{\rm str}- \nabla P^{\epsilon}_{\rm str} &= -\begin{bmatrix}
     \gamma_x \frac{\partial}{\partial z} \\
     \gamma_y \frac{\partial}{\partial z} \\
     \gamma_x \frac{\partial }{\partial x} +\gamma_y \frac{\partial}{\partial y} 
     \end{bmatrix} \frac{15 \epsilon^4}{8\pi \left(r^2 + \epsilon^2 \right)^{\frac{7}{2}}} \nonumber  \\
     \nabla \cdot  \mathbf{U}_{\rm str}^{\epsilon}&=0 \nonumber 
\end{align}
from which we derive (see \ref{sec:regStresslet}):
\begin{align}
    \mathbf{U}_{\rm str}^{\epsilon} &=  -\frac{5}{4  (\epsilon^2 + r^2 )^{5/2}}  \begin{bmatrix}
            2  xz(\gamma_x x + \gamma_y y ) + \epsilon^2 \gamma_x z \\
            2 yz(\gamma_x x + \gamma_y y ) + \epsilon^2 \gamma_y z \\
            2  z^2(\gamma_x x + \gamma_y y ) + \epsilon^2 (\gamma_x x + \gamma_y y) 
            \end{bmatrix}
\end{align}
After regularizing the Dirac delta function, Eqn. \ref{eq:leadingOuter} is recast in terms of $\mathbf{V}_{\rm ns} = \mathbf{U}_0^{(1)}-\mathbf{U}_{\rm str}^{\epsilon}$ and $P_{\rm ns} = P_0^{(1)} - P_{\rm str}^\epsilon$:
   \begin{align} \label{eq:nonSingular}
     \Delta \mathbf{V}_{\rm ns} {-}\nabla P_{\rm ns} {-} \mathbf{V_{ns}} {\cdot} \nabla \mathbf{\bar{U}}' {-} \mathbf{\bar{U}}'{\cdot} \nabla \mathbf{V_{ns}} &=    \mathbf{\bar{U}}' \cdot \nabla \mathbf{U}_{\rm str}^\epsilon {+} \mathbf{U}_{\rm str}^\epsilon {\cdot} \nabla \mathbf{\bar{U}}'  \\
     \nabla {\cdot} \mathbf{V_{ns}} &= \mathbf{0} \nonumber\\
     \mathbf{V_{ns}} &= -\mathbf{U}_{\rm str}^\epsilon ~\text{on channel walls}~. \nonumber 
 \end{align}
 
\section{Numerical Method}
  Equations \ref{eq:CTS},\ref{eq:singular},\ref{eq:nonSingular} are derived from different regularizations of Eqn. \ref{eq:leadingOuter}. To prevent the scaling of our computational domain from depending on particle Reynolds numbers, we introduce new dimensionless variables with lengths non-dimensionalized by the channel length scale: $\mathbf{x} = \mathbf{x}^{**} L$, pressures as: $p = \frac{\mu U_{\rm max}}{L} p^{**}$, and velocities as $\mathbf{V} = U_{\rm max}\mathbf{V}^{**}$. All three equations then take a common form:
 \begin{align}
     Re_c\left(\mathbf{\bar{U}}^{'**}\cdot \nabla \mathbf{V}^{**} + \mathbf{V}^{**} \cdot \nabla \mathbf{\bar{U}}^{'**}\right)  - \Delta \mathbf{V}^{**} +\nabla Q^{**} &=  \mathbf{G} \label{eq:commonform}\\
     \nabla \cdot \mathbf{V}^{**} &= 0 \nonumber
 \end{align}
 in which the body force, $\mathbf{G}$, depends on the type of regularization chosen and the field $Q^{**}$ stands in for $-P_D$, $-P_0^{(1)}$, or $-P_{\rm ns}$ respectively. In all three systems, $\mathbf{V}$ satisfies Dirichlet boundary conditions on the channel walls. Scaling lengths by $L$ means that channel geometries no longer depend on particle size and speed, at the cost of reintroducing $Re_c$ as a parameter in the PDE. In all formulations we must first solve for the background flow: $\mathbf{\bar{U}}^{**}\equiv (0,0,\bar{U}^{**}(x,y))$. Because the flow is rectilinear, $\bar{U}^{**}$ satisfies a 2D Poisson equation, which can be solved one time, for all particle locations and values of $Re_c$.
 
 Returning to Eqn. \ref{eq:commonform}, we simplify our solution geometry by taking a Fourier transform $z\rightarrow k$, at the same time omitting the double asterisks that denote dimensionless variables:
 \begin{align} 
 Re_c(-ik\bar{U}'\hat{V}_x) - \frac{\partial^2 \hat{V}_x}{\partial x^2} - \frac{\partial^2 \hat{V}_x}{\partial y^2} + k^2\hat{V}_x  + \frac{\partial \hat{Q}}{\partial x} &= \hat{G}_x \nonumber \\
  Re_c(-ik\bar{U}'\hat{V_y}) - \frac{\partial^2 \hat{ V_y}}{\partial x^2} - \frac{\partial^2 \hat{V}_y}{\partial y^2} + k^2\hat{V}_y   + \frac{\partial \hat{Q}}{\partial y} &= \hat{G}_y \label{eq:TransformedEQNS}
 \\
   Re_c(\frac{\partial \bar{U}'}{\partial x} \hat{V}_x + \frac{\partial \bar{U}'}{\partial y}\hat{V}_y) - \frac{\partial^2 \hat{ V_z}}{\partial x^2} - \frac{\partial^2 \hat{V}_z}{\partial y^2} + k^2\hat{V}_z  -ik\hat{Q}  &= \hat{G}_z 
   \nonumber \\
   \frac{\partial \hat{V}_x}{\partial x} + \frac{\partial \hat{V}_y}{\partial y} - ik\hat{V}_z &= 0 \nonumber
 \end{align}
 thereby reducing our equations to set of uncoupled 2D-PDEs for each of the Fourier modes. The 2D-PDEs for different wavenumbers can be solved in parallel. 
 
 We discretize our equations in $x$ and $y$ using the Taylor-Hood P2+P1 finite element method, in which the velocity is approximated using piecewise quadratic functions, $\phi_j$, and the pressure is approximated using piecewise linear functions, $\psi_j$ \cite{taylor1973numerical}:
 \begin{align}
 \hat{V}_i = \sum_j U_{ij}\phi_j &&  \hat{Q} = \sum_j P_{j}\psi_j
 \end{align}
 and then we integrate both sides of Eqn. \ref{eq:TransformedEQNS} against test functions $\phi$ for the momentum equations and $\psi$ for the divergence-free condition to obtain the discrete approximation of Eqn. \ref{eq:TransformedEQNS}:
 \begin{align}
     \begin{bmatrix}
     \mathbf{F} && \mathbf{B}^T \\
      \mathbf{B} &&  \mathbf{C}
     \end{bmatrix}
     \begin{bmatrix}
     \mathbf{U} \\
      \mathbf{P}
     \end{bmatrix}
     &=
     \begin{bmatrix}
     \mathbf{G} \\
      \mathbf{0}
     \end{bmatrix} \label{eq:matrixeqn}
 \end{align}
 where $\mathbf{F}$ is our discrete integral of the momentum equations and $\mathbf{B}$ is the discrete integral approximation of $-1$ times the divergence operator. Strictly, $\mathbf{C}=\mathbf{0}$. However, the resulting system has a non-trivial kernel containing all solutions with $\mathbf{V}=\mathbf{0}$ everywhere, and $Q=$ constant. To remove this degeneracy, we define $\mathbf{C}=\mathbf{0}$ for $k\not=0$ and for $k=0$ define $\mathbf{C} = c\mathbf{1}$, with $c$ chosen so as not to change the spectra of the matrix: $\lambda_{min}(\mathbf{F})<c<\lambda_{max}(\mathbf{F}) $. Each wavenumber, $k$, produces a different matrix on the left-hand side of Eqn. \ref{eq:matrixeqn}; however the matrix equations are decoupled and can be solved in parallel. 
 
 Once all coefficients $\mathbf{U}$ and $\mathbf{P}$ are solved for, we reconstitute $\mathbf{V}$ and evaluate it at the particle location to obtain the inertial focusing velocity. To identify focusing positions we need to find inertial focusing velocities for many candidate particle positions, which is equivalent to solving Eqn. \ref{eq:matrixeqn} with different forcing vectors $\mathbf{G}$ on the right-hand side but the matrices on the left-hand side that differ only by a Galilean transformation based on the particle's velocity.  This transformation is small enough that the same preconditioner is effective regardless of particle velocity.
 
 \subsection{Preconditioning}
 Eqn. \ref{eq:matrixeqn} is a large and sparse system of equations which, with an appropriately chosen preconditioner, can be solved using an iterative preconditioned Krylov subspace method (GMRES) that is more efficient and has less of a memory footprint than a direct method of solution based upon matrix factorization techniques, e.g. LU or QR factorization. An ideal right preconditioner for our matrix is
 \begin{align}
     \begin{bmatrix}
     \mathbf{F} && \mathbf{B}^T \\
      \mathbf{0} && - \mathbf{S}
     \end{bmatrix}^{-1} \label{eq:idealpreconditioner}
 \end{align}
 where $\mathbf{S} =  \mathbf{C}- \mathbf{B} \mathbf{F}^{-1} \mathbf{B}^T$ is the Schur complement. We call this preconditioner ``ideal'' because it will converge in just two iterations for all values of $\mathbf{G}$. To reduce the cost of applying this preconditioner, we can factor it as: 
 \begin{align}        \begin{bmatrix}
     \mathbf{F} && \mathbf{B}^T \\
     \mathbf{0} && -\mathbf{S}
     \end{bmatrix}^{-1} =
    \begin{bmatrix}
     \mathbf{F}^{-1} && \mathbf{0} \\
     \mathbf{0} && \mathbf{I}
     \end{bmatrix}
     \begin{bmatrix}
     \mathbf{I} && -\mathbf{B}^T \\
     \mathbf{0} && \mathbf{I}
     \end{bmatrix}
     \begin{bmatrix}
     \mathbf{I} && \mathbf{0} \\
     \mathbf{0} && -\mathbf{S}^{-1}
     \end{bmatrix} ~.
 \end{align}
 However the cost of applying this preconditioner is more expensive than solving Eqn. \ref{eq:matrixeqn} by a direct method, since it entails finding $\mathbf{F}^{-1}$ and $\mathbf{S}^{-1}$. We therefore construct an approximation of Eqn. \ref{eq:idealpreconditioner} from good approximations for $\mathbf{S}$ and $\mathbf{F}$ that are easy to invert. An approximation of $\mathbf{S}$ is based on the idea that if we replace our discrete operators, which do not commute, with continuous ones, which do, we can perform algebraic manipulations that will simplify the construction of the preconditioner \cite{wathen2003new}. We approximate $\mathbf{F}\approx \mathbf{\tilde{F}} = -\Delta$ and:
 \begin{align}
      \mathbf{S} &=  \mathbf{C} - \mathbf{B} \mathbf{F}^{-1} \mathbf{B}^T  \nonumber \\
     &\approx  \mathbf{C} -\nabla \cdot (\tilde{\mathbf{F}})^{-1}\nabla \nonumber \\
     &=  \mathbf{C}-(-\Delta )^{-1}\Delta \\
     &= \mathbf{C} + \mathbf{I}. \label{eq:continuousapproxS}
 \end{align}
  Represented in the FE basis this operator becomes:
 \begin{align}
     S_{ij}\approx \tilde{S}_{ij}  =\mathbf{C} + \int_{\Omega}\psi_i\psi_j\,d\mathbf{x}.
     \label{eq:tildeSdef}
 \end{align}
Hence our approximate preconditioner is
\begin{align}
  \begin{bmatrix}
     \mathbf{F} && \mathbf{B}^T \\
     \mathbf{0} && - \mathbf{S}
     \end{bmatrix} \approx
       \begin{bmatrix}
     \mathbf{\tilde{F}} && \mathbf{B}^T \\
     \mathbf{0} && - \mathbf{\tilde{S}}
     \end{bmatrix}
     \end{align}
     where $\mathbf{\tilde{F}} = -\Delta$ and $\mathbf{\tilde{S}}$ is as defined in  Eqn. \ref{eq:tildeSdef}. We factor the application of this preconditioner as
\begin{align}
         \begin{bmatrix}
     \mathbf{\tilde{F}} && \mathbf{B}^T \\
     \mathbf{0} && - \mathbf{\tilde{S}}
     \end{bmatrix}^{-1} =
    \begin{bmatrix}
     \mathbf{\tilde{F}}^{-1} &&  \mathbf{0} \\
      \mathbf{0} &&  \mathbf{I}
     \end{bmatrix}
     \begin{bmatrix}
      \mathbf{I} && -\mathbf{B}^T \\
      \mathbf{0} &&  \mathbf{I}
     \end{bmatrix}
     \begin{bmatrix}
      \mathbf{I} &&  \mathbf{0} \\
      \mathbf{0} &&  -\mathbf{\tilde{S}}^{-1}
     \end{bmatrix}. 
 \end{align}
 This makes the cost of preconditioning one application of $\mathbf{\tilde{F}}^{-1}$, one application of $\mathbf{\tilde{S}}^{-1}$, and one application of $\mathbf{B}^T$.  
 
 Calculating particle focusing positions for one channel involves tens of thousands of solves of Eqn. \ref{eq:matrixeqn} for different wavenumbers and particle locations (Fig. \ref{fig:squareChannelFigure}). As each solve is independent, we parallelize over wavenumbers and assign each worker solving Eqn. \ref{eq:matrixeqn} a single wavenumber but for all sampled particle locations. The worker constructs the preconditioner for this wavenumber along with its column- and row- pivoted LU factorization, and uses these to solve the matrix equation for every particle location. Column- and row- pivoting accelerates the LU factorization, and reduces the number of nonzero elements that need to be stored, thereby speeding up preconditioning.  In our tests, GMRES converged with a relative residual of $10^{-4}$ in between 4 and 15 iterations for all values of $k$ for $Re_c=1$ and between 20 and 100 iterations for $Re_c=100$.

 \subsection{Integration Of The Right Hand Side}
\label{sec:RHS}
$\mathbf{V}$ can be accurately approximated on a coarse finite element mesh because it is continuous, however the right hand side of Eqn. \ref{eq:matrixeqn} diverges like $O(r^{-1})$ at the location of the particle, meaning that integrals of the right hand side are seldom accurate when approximated on the same coarse grid. To improve the accuracy, the integrals over the element containing the singularity and its $D$ nearest neighboring elements were obtained by combining the integrals over a $4^n$ congruent sub-triangulation of the elements. The integrals over all triangles and sub-triangles were
obtained using second order Gauss quadrature. Our experiments showed $D=1$ and $n=2$ were sufficient to accurately integrate the right-hand side and were used for all further calculations.
 
\subsection{Calculating a Particle Trajectory} 
\label{sec:trajectory}
To calculate the trajectory of a particle across the channel cross-section, we must solve the ordinary differential equation:
\begin{align}
\label{eq:ode}
    \frac{d}{dt}(\mathbf{x}_p) = \mathbf{V}(\mathbf{x}_p;\mathbf{x}_p)
\end{align}
where $\mathbf{V}(\mathbf{x}_p;\mathbf{x}_p)$ denotes evaluating the solution of Eqn. \ref{eq:leadingOuter} at $\mathbf{x}_p$ with the singularity located at position $\mathbf{x}_p$. Because the singularity must be at $\mathbf{x}_p$, the RHS of Eqn. \ref{eq:leadingOuter} is different for every particle location, meaning one PDE solve is required for each time step.

\begin{figure}
 \centering
     \includegraphics[width =\textwidth]{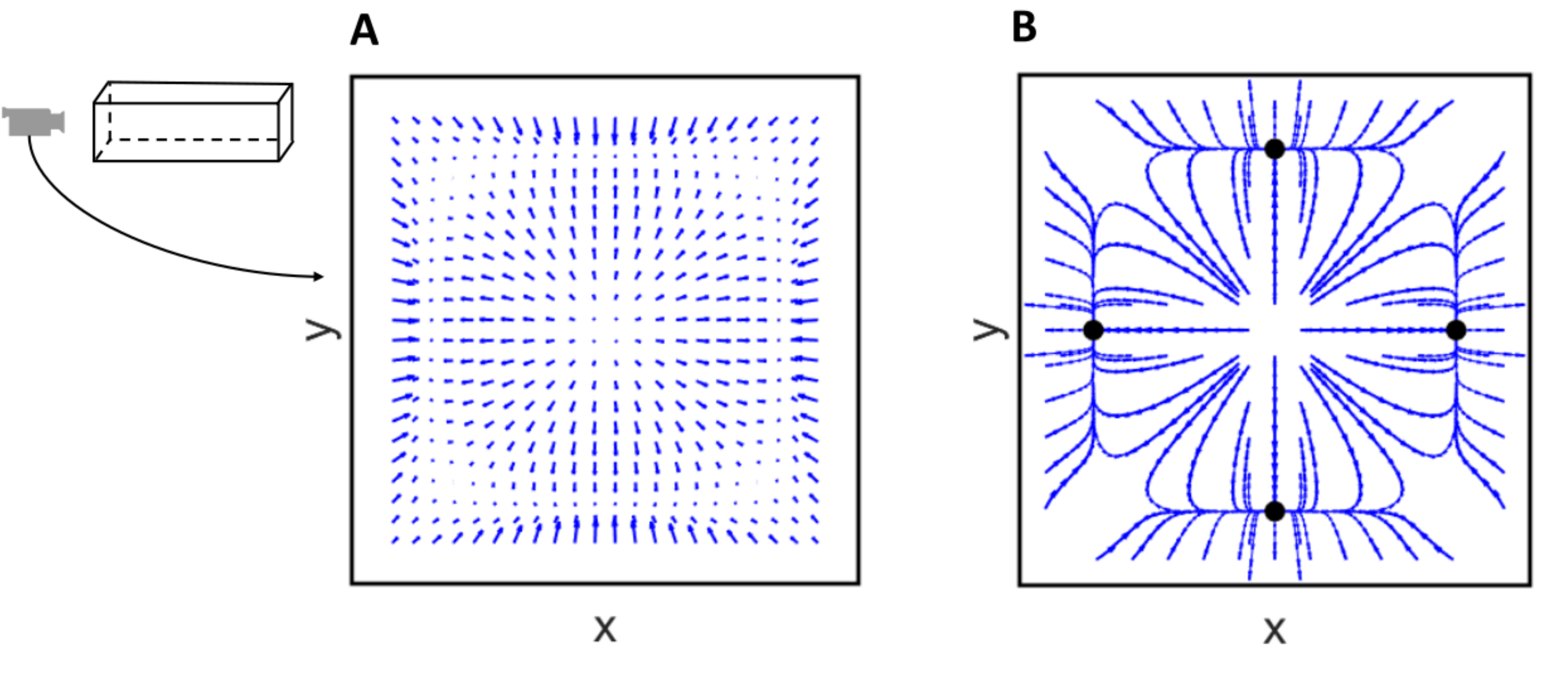}
     \caption{Migration velocities are sampled at a grid of particle locations, and natural neighbor interpolation is used to evaluate velocities across the entire channel cross-section. All views show only the $x-$ and $y-$ components of the migration velocity. A: Migration velocities are sampled on a grid spaced by $0.03$ in $x$ and $y$.
     B: The sampled migration velocities are then interpolated to reconstruct particle trajectories by solving Eqn. \ref{eq:ode} using ode45, and stable focusing positions are found as the limit points of trajectories (black dots).}
     \label{fig:squareChannelFigure}
 \end{figure}

Our goal is to calculate trajectories with initial conditions covering the channel, therefore it is more efficient to pre-calculate the migration velocity on a grid of sample positions (Fig. \ref{fig:squareChannelFigure}A) and interpolate between them to calculate $\mathbf{V}(\mathbf{x}_p;\mathbf{x}_p)$ in order to advance the particle trajectory. Once velocities are sampled across the entire channel cross-section, we advect the particles by solving Eqn \ref{eq:ode} using the fourth order Runge-Kutta solver, ode45 in MATLAB (Mathworks, Natick, MA). We calculate the trajectories of many particles, identify stable focusing positions by considering the limiting behaviors of the particles (Fig. \ref{fig:squareChannelFigure}B).
 
In order to handle arbitrary channel shapes, we use natural neighbor interpolation\cite{sibson1981brief} to interpolate velocities between arbitrarily spaced sampling points. This interpolation is once differentiable everywhere except the sampling points and yields quadratic convergence upon refinement.


\section{Results}

\subsection{Numerical Convergence}
\begin{figure}[ht]
     \centering
     \includegraphics[width=\textwidth]{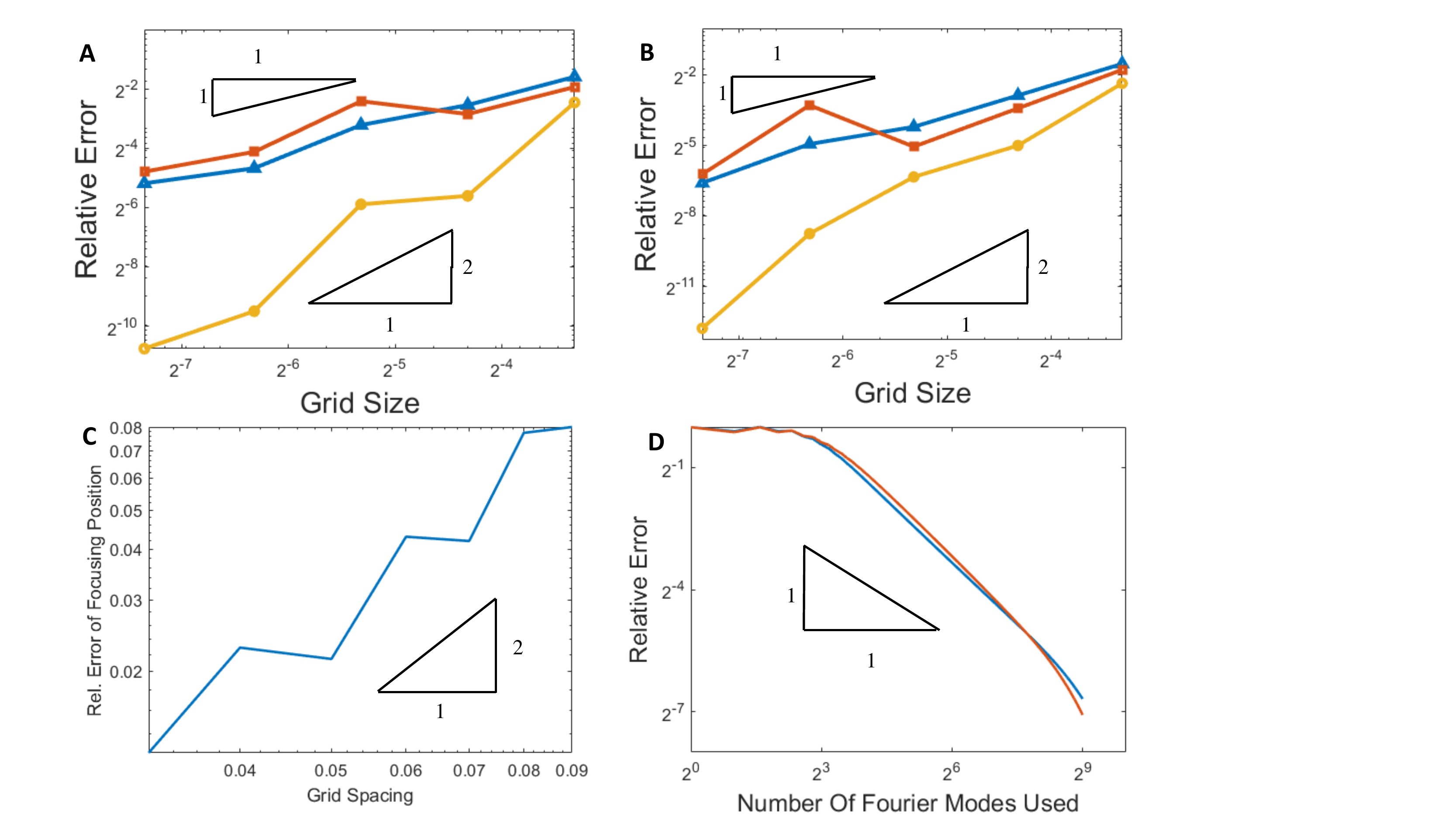}
     \caption{Testing convergence of the inertial migration velocity with respect to mesh density at $R_c=1,\ 50$ (A\&B), sampling point density (C), and Fourier modes (D).  A\&B: Relative error of migration velocity at $R_c=1$(A) and $R_c=50$(B) of the 3 different methods of solution Eqn. \ref{eq:CTS} in yellow, Eqn. \ref{eq:singular} in blue, and Eqn. \ref{eq:nonSingular} in red. C: Relative error of migration velocity versus number of Fourier modes included for $R_c=1$ (blue) and $R_c=50$ (red).  D: Convergence of focusing positions (sec. \ref{sec:FocusingPositions}) as we increase sampling point density with $R_c=50$ and a channel cross section of scalene triangle with side lengths 1, 1.28, 1.02.  A scalene triangle was chosen so that symmetry does affect convergence.}
     \label{fig:ElemConv}
 \end{figure}

In this section we demonstrate the accuracy and efficiency of the migration velocity calculation and validate its second order convergence. To test our regularization method from Eqn. \ref{eq:CTS}, we perform a self-convergence test on the calculated migration velocity for a particle in a $1\times 1\times 10$ square channel, with one particle located at $\mathbf{x}_p=[-0.1,0.2]$ and $Re_c = 1,\ 50$. We also compare results and convergence rates  the regularization method of Eqn. \ref{eq:CTS} with the existing regularization methods given by Eqns. \ref{eq:singular} and \ref{eq:nonSingular}(Fig \ref{fig:ElemConv}A\&B).

\emph{Mesh convergence for different regularization methods}. We measure the rate of convergence for our migration velocity calculation as we decrease the 2D mesh size in a 0.1x0.1 box around the particle.  Convergence is measured by comparing to computations done with half the grid size of the smallest grid size plotted. Relative error decreases proportional to $O(h^2)$ as the local grid-size around the discontinuity/singularity is reduced (Fig. $\ref{fig:ElemConv}$A\&B), which is the theoretical optimal rate of convergence for the $P2+P1$ scheme.  

We compare our new regularization method with two existing techniques, which were described in section \ref{sec:alternativeFormulations}. Both existing methods regularize the particle-singularity by blunting either the Dirac delta function (Eqn. \ref{eq:singular}) or the stresslet velocity (Eqn. \ref{eq:nonSingular}), and introduce a blunting length-scale, $\epsilon$.  Our goal is to choose this length scale so as not to lose the physics induced by the singularity, while still representing it with our finite element basis. Empirically, we find minimum values for $\epsilon$: $\epsilon = \textrm{gridsize}/2 $ for Eqn. \ref{eq:singular} and $\epsilon = \textrm{gridsize}/4$ for Eqn. \ref{eq:nonSingular}.
 The existing methods had almost identical accuracy and $O(h)$ rates of convergence (Fig. \ref{fig:ElemConv}A), and at the smallest mesh-sizes considered, produced errors that were around 16-fold larger than Eqn. \ref{eq:CTS}. The slower convergence of the alternate methods is inevitable, since each makes an $O(\epsilon)$ approximation of the particle singularity. Indeed, even after the singularity is removed the velocity field remains discontinuous at the particle location. The remainder of the computational results presented utilize the method based upon Eqn. \ref{eq:CTS}

 
\emph{Convergence in number of velocity sampling points} 
Rather than evaluating the migration velocity in Eqn. \ref{eq:ode} directly, we pre-evaluate velocities on a grid of sampling points, and use interpolation to compute velocities at intermediate points. Our velocity interpolation method gives second order convergence as the spacing of sampling points is reduced \cite{sibson1981brief} and the results from Fig. \ref{fig:ElemConv}C shows this achieved for focusing position convergence. Fairly coarse sampling meshes can be used over much of the channel cross-section, however reconstructing the slow focusing manifold and the boundaries between the basins of attraction requires locally fine sampling meshes. Since the locations of these curves is not known \emph{a priori}, we default to using a sampling mesh of 0.04.  

\emph{Convergence in Fourier modes.} The number of Fourier modes used to represent the solution is another parameter that influences the accuracy of the method. In Fig. \ref{fig:ElemConv}D the convergence behavior is shown as an increasing number of Fourier modes are use to resolve the $z-$dependence of the particle disturbance flow. Errors in migration velocity decay like $O(N^{-1})$, where $N$ is the number of Fourier modes. Convergence in Fourier modes is not spectral, likely because continuity of the velocity fields we are solving for fails at the level of the first derivative. Convergence is unaffected when changing $R_c$ from 1 to 50. 

\subsection{Calculating Focusing Positions In Channels of Arbitrary Cross Section}
\label{sec:FocusingPositions}
\begin{figure}[ht]
     \centering
     \includegraphics[width=\textwidth]{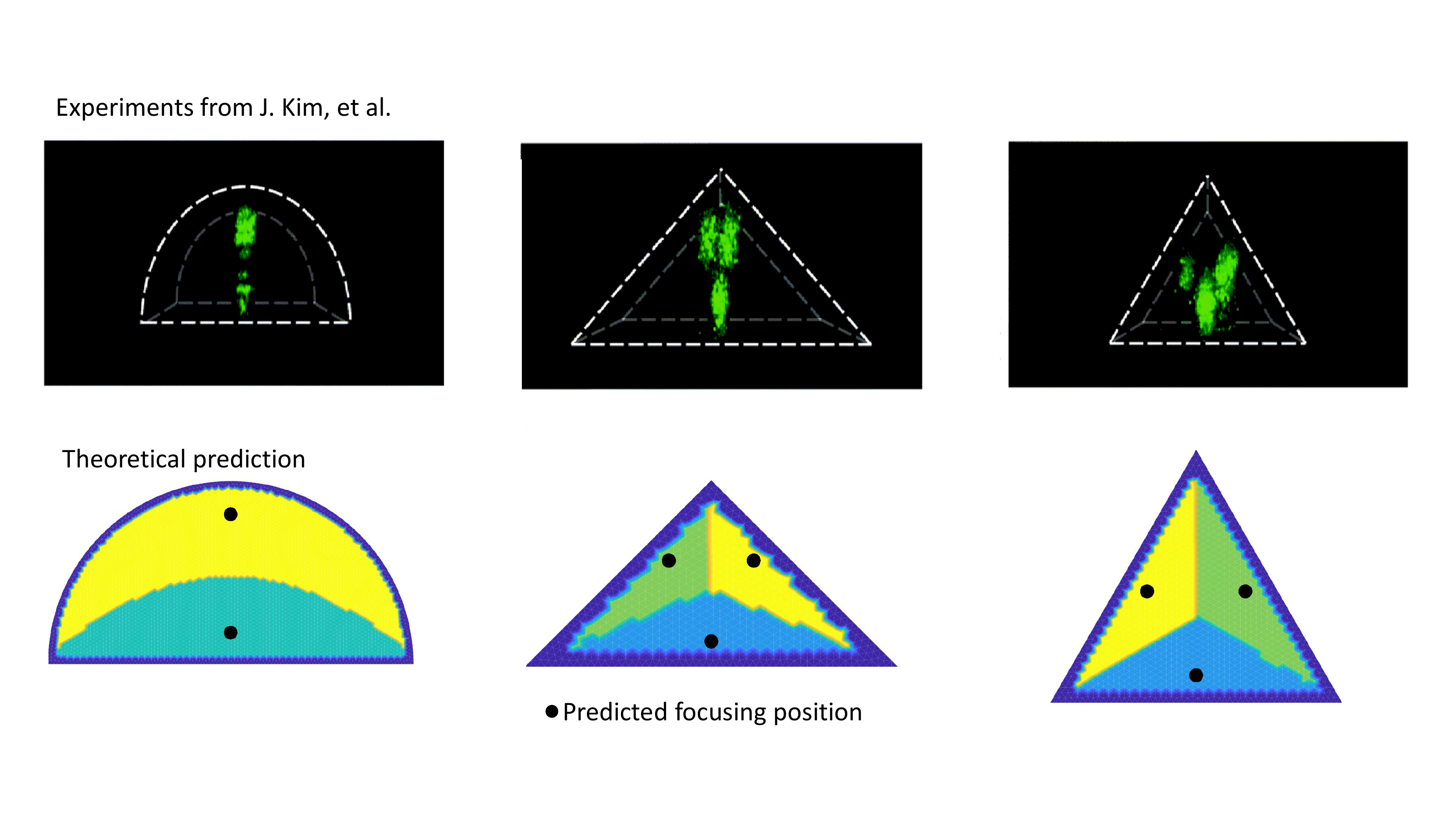}
     \caption{Predicted focusing positions from our fast simulation method agree well with experimental data from \cite{kim2016inertial}. From left to right: particle focusing positions in channels with semicircle, 2x1 isosceles triangle, and equilateral triangle cross section. Reynolds numbers were matched with those from experiment.}
     \label{fig:geometryComparison}
 \end{figure}

Inertial migration causes particles to eventually collect at equilibrium streamlines, known as \textbf{focusing positions} (Fig. \ref{fig:squareChannelFigure}). We find stable focusing positions by integrating Eqn. \ref{eq:ode} until particle velocities decrease below $10^{-5}$. We identify all focusing positions by seeding the channel cross-section evenly with thousands of particles. We color code the seeding location according to which focusing position the particle converges to.  This process partitions the cross section into basins of attraction (Fig. \ref{fig:geometryComparison}).

In a square channel, in common with previous studies \cite{hood_lee_roper_2015}, we find four symmetrically arranged focusing positions, one at each mid-face (Fig. \ref{fig:squareChannelFigure}B). The slow focusing manifold symmetrically links all four stable focusing positions. Although not shown explicitly, we infer that there must be four symmetrically arranged unstable focusing positions -- that is streamlines that particles may remain on, but can not converge to -- where the slow focusing manifold passes through the four diagonals of the square. Omitting these four unstable positions and the unstable center, the basins of attraction divide the channel into four triangular quadrants.
 We further validate predictions about focusing positions by comparing our computations with real experiments \cite{kim2016inertial} using microfluidic channels with three different, non-rectangular cross-section shapes: two triangles of different base-height ratios and a semi-circle. We match $Re_c$ to the parameters in \cite{kim2016inertial}. Although no experimental data was presented on the trajectories of particles, we found close agreement between predicted and experimentally measured final focusing positions. In particular we see the same number of focusing positions (2 in a semicircular channel, and 3 in a triangular channel), as well as narrowing of the space between focusing positions in the larger base-height ratio triangle.
\begin{figure}
    \centering
   \includegraphics[width = 0.95\textwidth]{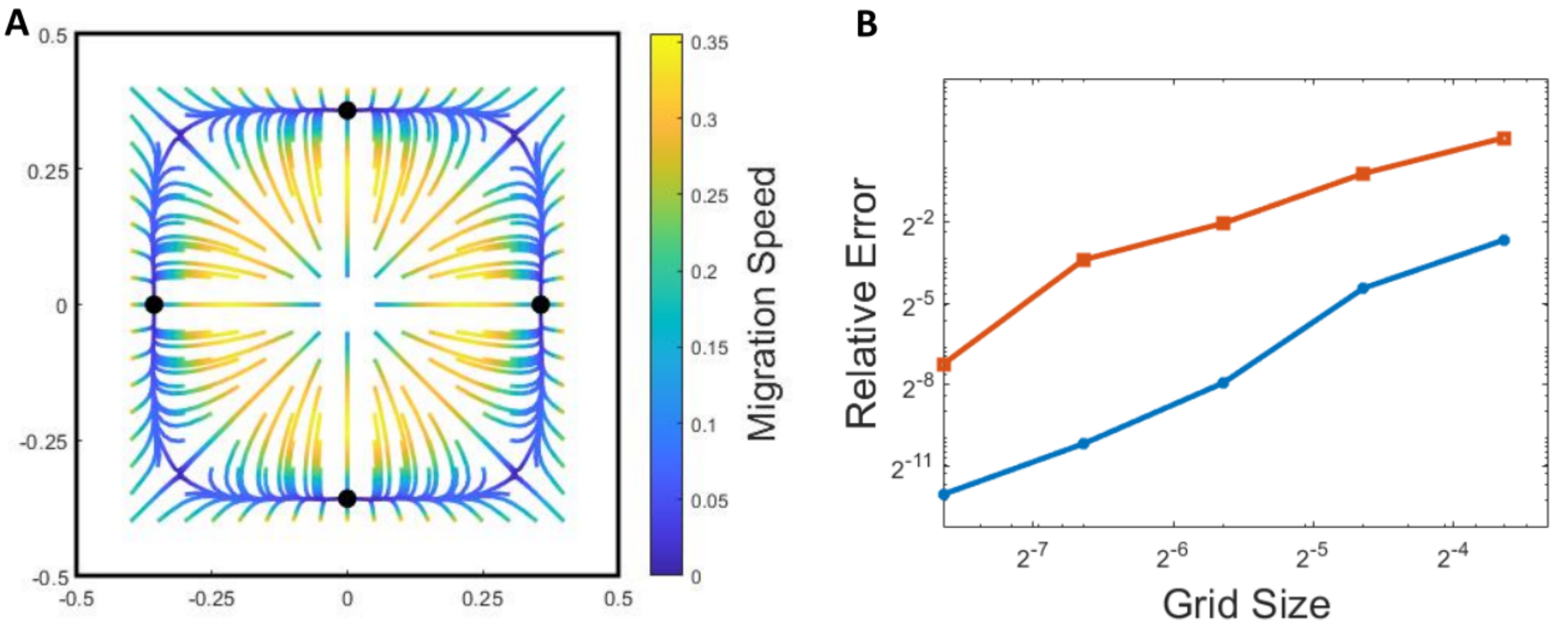}
    \caption{Particles quickly migrate away from the center of the channel towards the slow focusing manifold, where they slowly migrate to the focusing positions. A: Particle trajectories are plotted with color denoting migration speed at a point. The black dots represent the focusing positions, the streamlines all particles will eventually gather upon.  B: Computed migration velocities converge as $O(h^2)$ both on the slow manifold (red squares) and away from the slow manifold (blue circles). However, relative error is much larger on the slow manifold because velocities are ten-fold smaller.  Tests were done in a square channel with $R_c=1$ at $\mathbf{x}_p=(-0.2,-0.1)$ (far from the slow manifold) and $\mathbf{x}_p=(-0.1,-0.355)$ (close to the slow manifold).}
    \label{fig:slowVsFastConvergence}
\end{figure}

\subsection{Convergence Along Fast and Slow Manifolds} Inertial focusing occurs in two phases (Fig. \ref{fig:slowVsFastConvergence}) \cite{hood2016direct}: first particles move rapidly from their starting streamline toward a heteroclinic orbit within the channel cross-section (i.e. a manifold of streamlines when the stream-wise dimension is considered). Particles then move much more slowly along this manifold. Indeed, experimental evidence suggests that in real inertial microfluidic devices particles perform much of their fast focusing in the device inlet, leaving only the slow focusing parts of trajectories to be resolved within the channel \cite{hood2016direct}. Resolving the fast and slow dynamics is very important to understanding the speed at which particles migrate towards their focusing positions, as well as the size of the region of attraction.

Accurately calculating the migration velocity is more difficult along the slow manifold than elsewhere in the channel.  This is because the magnitude of the error is similar but the migration velocity is $\sim 10$ fold smaller, thus the relative error is much higher (Fig \ref{fig:slowVsFastConvergence}B). For example, we compare self convergence of the migration velocity in a square channel with $R_c=1$, between a particle located at $(x_p,y_p)=(-0.2,-0.1)$, far from the slow focusing manifold, with a particle located at $(x_p,y_p) = (-0.1,-0.355)$, which lies on the slow focusing manifold. Although both velocities converge asymptotically as $O(h^2)$, relative errors remain 10-fold larger close to the slow focusing manifold.

 As will be shown in Section \ref{sec:dynamics}, the location of the slow focusing manifold is sensitive to the physical parameters of the model; including channel shape and Reynolds number. The manifold shape also depends upon numerical parameters such as mesh size. Since whether a particle is placed on or off the slow manifold has a drastic effect upon the size of its inertial migration velocity, mesh size constraints are especially restrictive close to the slow manifold, since any perturbation of the manifold away from the particle will add a fast focusing velocity. For example, we compare self convergence of the migration velocity in a square channel with $R_c=1$, between a particle located at $(x_p,y_p)=(-0.2,-0.1)$, far from the slow focusing manifold, with $(x_p,y_p) = (-0.1,-0.355)$, which lies on the slow focusing manifold. Although both velocities converge asymptotically as $O(h^2)$, relative errors remain 10-fold larger close to the slow focusing manifold.
 Although variation in relative errors across the cross-section means that we need to be cautious in evaluating the convergence of our algorithm, we found that a mesh-size of .02 could be used to reliably distinguish trajectories associated with different Reynolds numbers and channel shapes. However, resolving dynamics on the slow manifold, such as bifurcations affecting the number of stable focusing positions (Section \ref{sec:dynamics}) is particularly challenging numerically. Close to the critical shapes or $Re_c$-values that produce these bifurcations, focusing becomes extremely slow, potentially leading to the appearance of supernumerary stable focusing positions \cite{chun2006inertial}.

 \subsection{Varying channel shape and Reynolds number} \label{sec:dynamics}

Our algorithm is fast enough to test many channel geometries and particle Reynolds numbers, allowing for rapid exploration of the effect of channel shape and flow Reynolds number upon the number and location of focusing positions; an area in which there are still many unanswered questions. 

\begin{figure}[ht] 
    \centering
    \hspace*{-0cm}
   \includegraphics[width=1.0\textwidth]{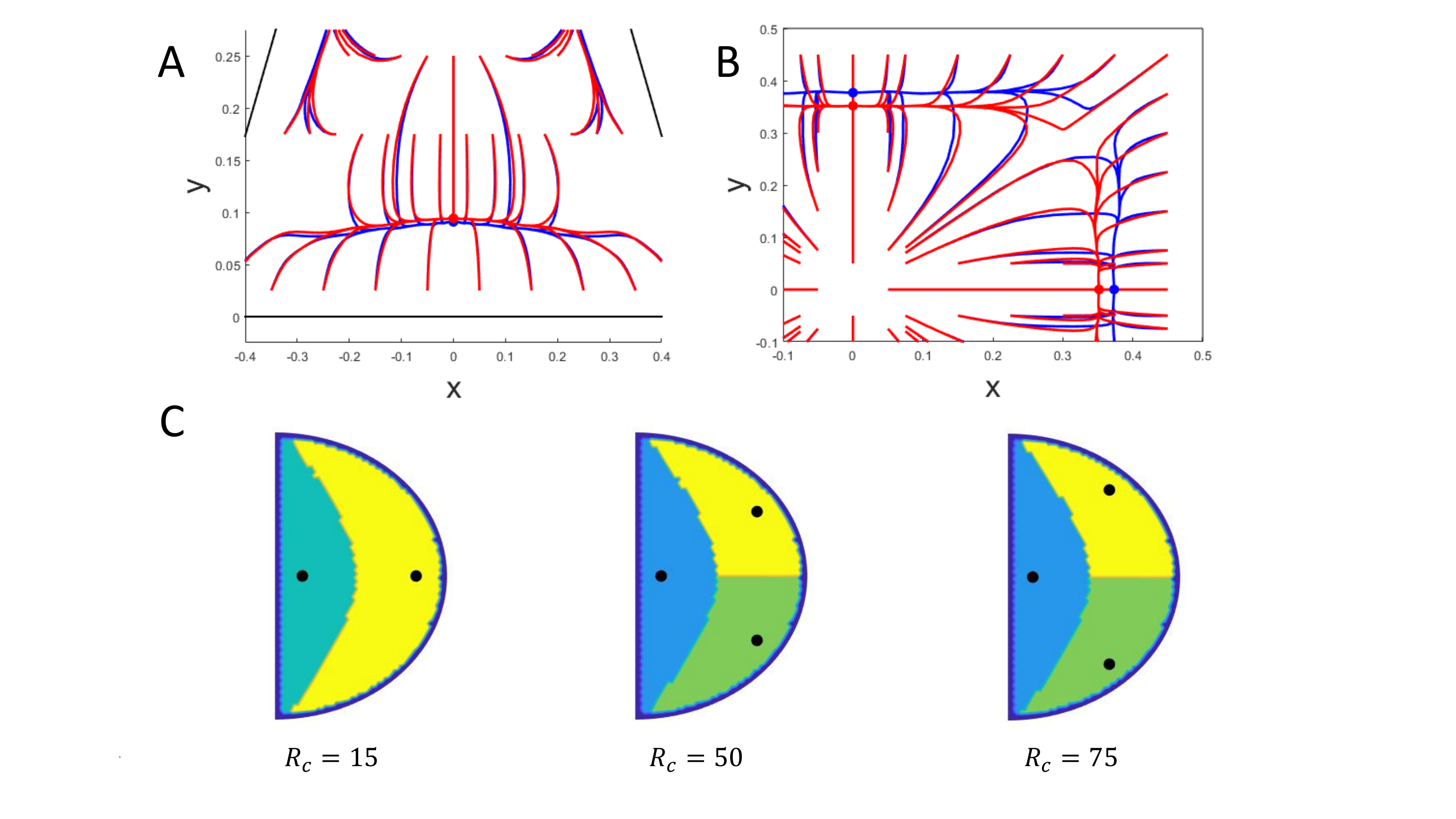}
      \caption{Increasing channel Reynolds number changes the position and number of focusing streamlines. A-B, $Re_c=1$ (red) and $Re_c=100$ trajectories in an equilateral triangle and a square. In the square, increasing $Re_c$ displaces focusing positions further from the center of the channel, but there is no effect in the triangle. C, Focusing positions bifurcate as $Re_c$ is increased in a semi-elliptical channel with aspect ratio 1.2.}
      \label{fig:REcomparison}
\end{figure}

As an example of the computational ability, we the effects of increasing Reynolds number in several geometries.  Our calculations in square channels support previous studies that show that the slow focusing manifold and focusing positions move toward the channel walls as $Re_c$ is increased (Fig. \ref{fig:REcomparison}B). Di Carlo et al. \cite{di2009particle} hypothesize that focusing positions arise from a balance of (shear) forces pushing the particle away from the channel center and (wall) forces that push particles away from channel walls, and that the shear forces increase more rapidly with $Re_c$. However, our modeling casts some doubt on this mechanism, because we find no such shift in focusing positions as $Re_c$ is increased for an equilateral triangle-shaped channel. In the equilateral triangle channel, focusing positions are symmetrically arranged near the middle of each channel wall, and maintain the same distance from the wall even as $Re_c$ is increased 100-fold (Fig. \ref{fig:REcomparison}A). 

Emerging methods for microfabrication now allow microfluidic channels to be built with ever more complex shapes, including channels with curved walls \cite{kim2016inertial}. Our computations show that the number of focusing positions undergoes a finite-$Re_c$ bifurcation in some curved channels. Inspired by \cite{kim2016inertial}, we simulated a semi-elliptical channel with major:minor axis aspect ratio 1.2. Migration velocity sampling points were taken on a square mesh with density 0.03. We found that for this channel there is an apparent supercritical pitchfork bifurcation at $Re_c\approx 40$ in which a single stable fixed point in the mid-point of the channel curved face splits into two stable fixed points, separated by an unstable fixed point (Fig. \ref{fig:REcomparison}C). The emergence of the fixed point by a pitchfork bifurcation stands in contrast to square channels, where, at $Re_c\approx 250$, new focusing positions emerge near the channel diagonals via \emph{saddle point} bifurcations \cite{yamashita2019bifurcation}. Close to the bifurcation, migration velocities along the slow manifold become extremely slow (`critical slowing down' \cite{strogatz2018nonlinear}), which increases the challenge of resolving the precise number of stable fixed points by studying the long time convergence of particles. Moreover, our model for inertial migration is only exact in the limit as $\alpha\to 0$ and $Re_p\to 0$; we hypothesize that perturbations associated with modeling finite-size particles could create the imperfect bifurcations \cite{strogatz2018nonlinear} seen in \cite{yamashita2019bifurcation}.

We also explore how number and location of stable focusing positions varies over the space of channel shapes. Because of the limitations of photo-lithographic methods, inertial microfluidic devices typically have rectangular channels. Experimental data seems to suggest that in contrast to square channels which have at least four stable focusing positions, channels with sufficiently high aspect (width to height) ratios, may have only two stable focusing positions, centered at the two larger edges of the channel \cite{di2007continuous,edd2008controlled,lee2010dynamic}. We simulated inertial focusing in 4 rectangular channels whose aspect ratios ranged from 1 to 4. Migration velocities were sampled on a rectangular mesh of points with spacing chosen to be a fraction $0.04$ of the channel size in each dimension, e.g. for 4$\times$1 rectangle the spacing was 0.16 in $x$ and 0.04 in $y$. 

\begin{figure}[ht]
    \centering
    \hspace*{-0cm}
    \includegraphics[width=1.0\textwidth]{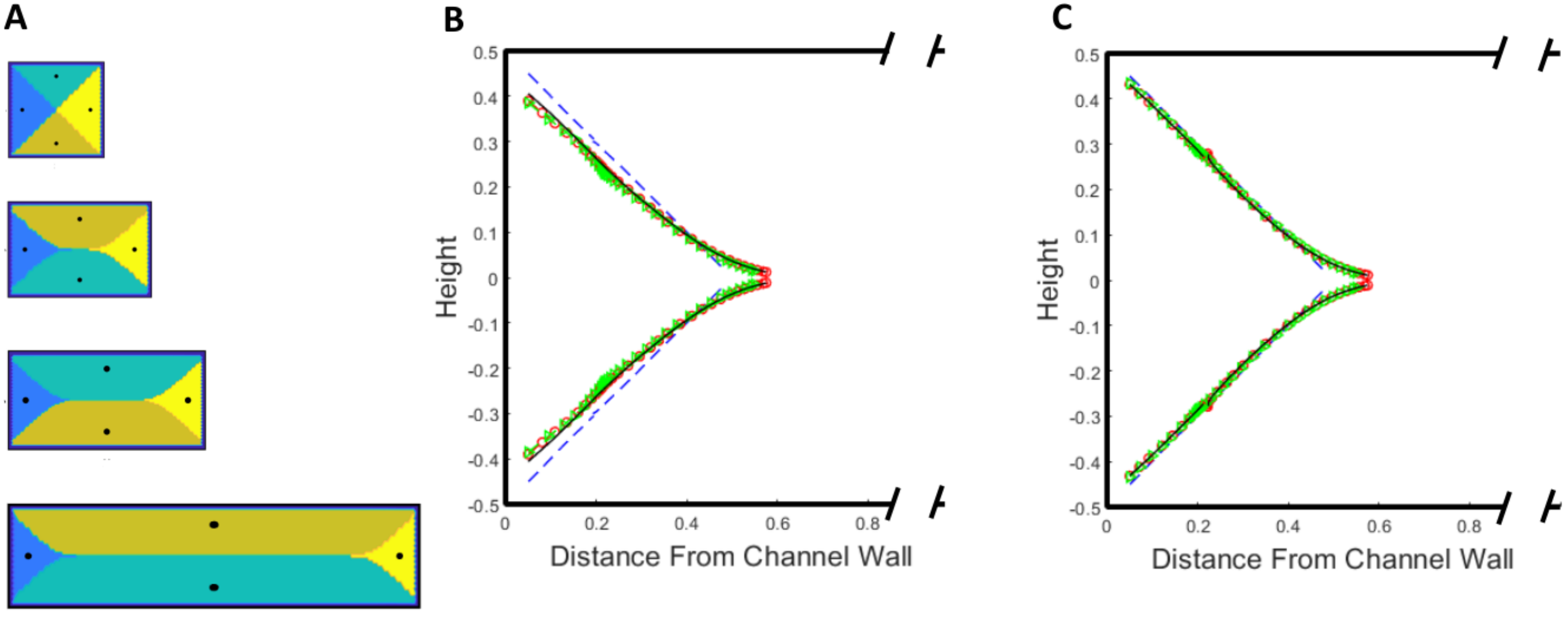}
    \caption{A sweep over a range of parameters shows that short-side focusing positions persist even in high aspect ratio channels. The basins of attraction of these focusing positions approach a limiting shape as aspect ratio is increased.  Left: Basins of attraction for channels of aspect ratio 1, 1.5, and 2. Right: Basins of attraction along the minor axis for rectangles of different aspect ratios.  The blue dashed line represents AR=1, while the red circles, green triangles, and solid black line represent AR = 1.5, 2, and 4. For all simulations $Re_c=1$. While the channels had different velocity sampling points (see main text), the basins of attraction created via interpolation were unaffected by sample spacing.}
    \label{fig:rectangleBA}
\end{figure}

We observe for low aspect ratio channels and consistent with experiments \cite{di2007continuous}, that channels have four stable focusing positions, each centered on one of the channel's four sides. However, even for the most `slot'-like channel geometries, our simulations showed particles continuing to converge to any of four stable stable focusing positions even in the high aspect ratio channels (Fig. \ref{fig:rectangleBA}, left panels), consistent with \cite{hood2016direct}. We studied the basins of attraction of the short-side focusing positions. We found that, rather than disappearing, the basins of attraction as the aspect ratio of the channel is increased, the basins of attraction for these short-side focusing positions approach a limiting shape as channel aspect ratio increases (the basins of attraction are shown for $Re_c = 1$ in Fig. \ref{fig:rectangleBA}, right panels). While an isthmus between the two major basins of attraction must exist, it quickly becomes too slender to resolve in our simulations. Since our method leverages a very different set of assumptions about channel Reynolds numbers to those of \cite{hood_lee_roper_2015}, our simulations again support the conclusion reached by \cite{hood2016direct}, that the disappearance of short-side focusing positions in experiments is an artifact of the prefocusing that occurs in the inlets that feed particles into real inertial microfluidic channels.

 \section{Conclusion} \label{sec:conclusion}

 By representing particles submerged in fluids by singularities Sch\"onberg and Hinch\cite{schonberg1989inertial} were able to accurately calculate particle behavior without solving the nonlinear Navier-Stokes equations or explicitly meshing the 3D and time-varying fluid domain between particles and channel walls. We have taken the asymptotic approximations created in \cite{schonberg1989inertial} and extended them to create a fast, efficient numerical method for calculating particle migration velocities in straight channels of uniform but arbitrary cross section. A proper accounting for singular and discontinuous terms that must be matched between inner and outer expansions allowed us to reduce the computation to one of finding only continuous components of the velocity field, improving the order of accuracy of the method and allowing for fast, stable calculations using a uniform mesh, including at the location of the particle itself. Numerical tests reveal that at moderate mesh sizes this method of regularizing the solution produces 10-fold higher accuracy than existing methods of blunting singularities, namely regularized stresslets and blob function approximations.


 The method can be extended to particles of other shapes. Singularity modeling of the particle needs only the stresslet strength associated with the Stokes solution around the particle. This stresslet strength is already known for ellipsoids in shear flow\cite{chwang1975hydromechanics} and for other particles can be found by solving for the motion of the particle in Stokes flow, for which there are many approximate or numerical methods \cite{happel2012low,kim2013microhydrodynamics}. However, a fully accelerated computation also requires that the discontinuous components of the velocity, $\mathbf{U}_D$, be found from the $O(\alpha R)$ inhomogeneous Stokes equation.  For spherical particles we were able to isolate and find these terms analytically but for more complex particle geometries, this regularization could be given to a specialized solver designed to solve the inhomogeneous Stokes equations.
 
 We note that subtracting the discontinuity leaves a velocity field that is continuous but that still has discontinuous derivatives. Higher order discontinuities can, in principle, be computed analytically by continuing the inner expansion to high order, that is, we can compute the discontinuities in the first derivative of $\mathbf{V}$ by continuing our expansion of the velocity field to $\mathbf{u}_2^{(1)}$, but numerical experiments in COMSOL Multiphysics (COMSOL Inc, Los Angeles, results not shown) showed that subtracting higher order discontinuities did not allow any reduction in the number of elements. Moreover, we found that the solver produces sufficiently accurate velocities at mesh sizes that are large enough to be applied uniformly across the entire of the channel cross-section, so discontinuity regularization alone is enough to achieve our goal of avoiding remeshing for different particle positions. However, the full potential of the Fourier basis used to represent the $z-$dependence of the solution remains unrealized.

Further expansions in terms of particle size and Reynolds number are possible, and would add either higher order forcing terms within our Oseen equation, or would require that we impose additional matching conditions to model the particle. Such an extension could be useful for predicting differential focusing of different particles within the same channel. Currently, although our simulations provide quick and accurate tools for predicting how particle focusing positions are affected by channel shape and channel Reynolds number, they can not reveal the effects of size upon focusing positions. In straight channels the effects of size are often mild \cite{kim2016inertial}, but particles of different sizes can follow very different focusing trajectories in non-straight channels, allowing particles of different sizes to be separated \cite{kuntaegowdanahalli2009inertial,mach2011automated}, and extending models to incorporate explicit dependence on particle size, offers a first step toward modeling the complex 3D geometries of a wide range of inertial microfluidic channels.  Hood et al. \cite{hood_lee_roper_2015} incorporated size-dependent focusing effects by calculating the next order term in $\alpha$: $\mathbf{u}_2^{(0)}$. Although our model relies on different asymptotic limits, it could be similarly extended, though the calculation would require generating new solutions in both inner and outer regions, but still of the form Eqn \ref{eq:commonform}, and thus we expect that the extension would not require any great increase of compute times.
 
\begin{figure}
     \centering
     \includegraphics[width=1.0\textwidth]{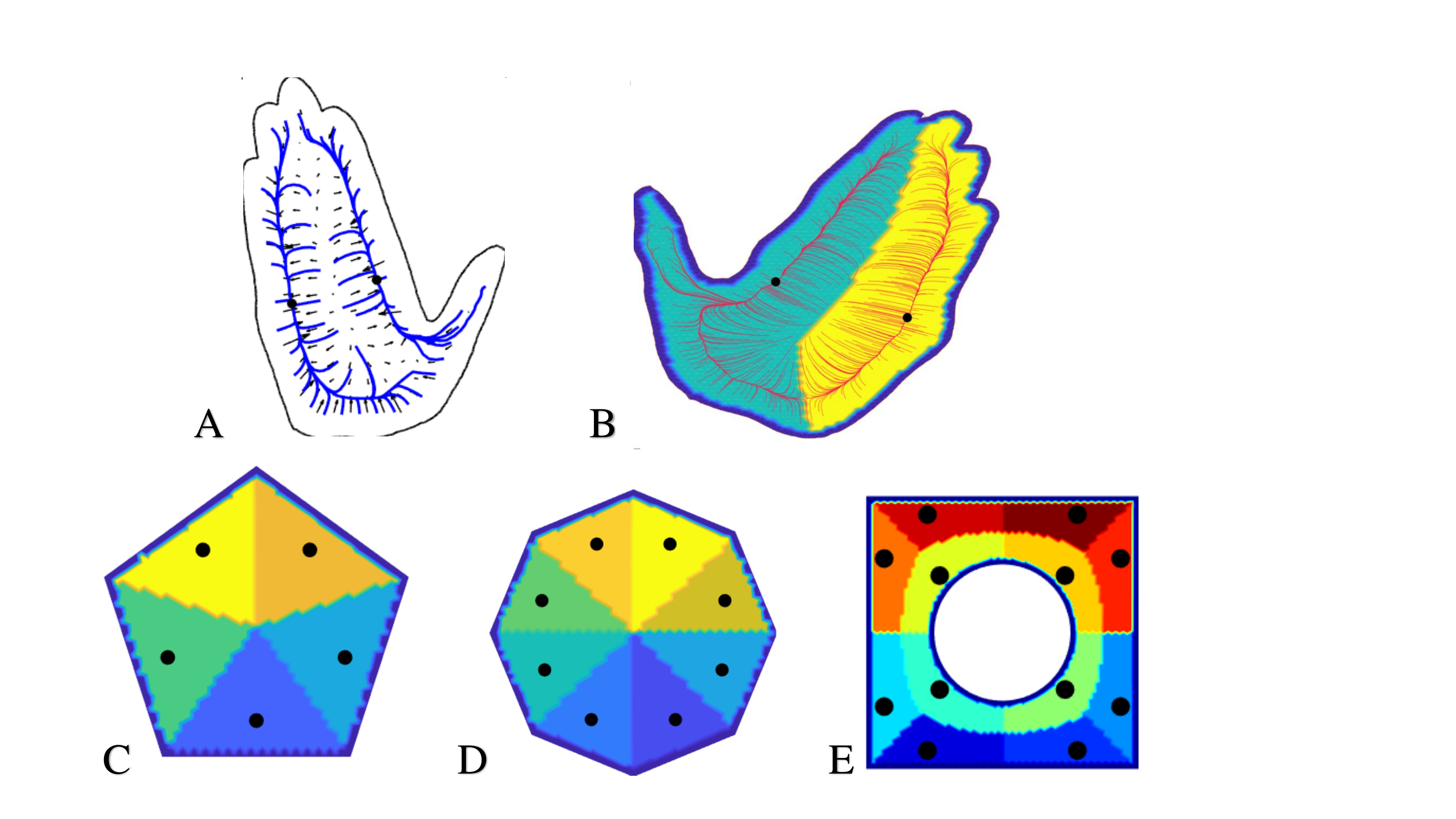}
     \caption{Memory costs and time for individual velocity solves are quite small ($\sim$200MB and $\sim$20 seconds on 1 core when averaged over simultaneous solves), allowing particle focusing velocities to be computed on a laptop computer. However, for complex channel geometries hundreds of particle positions are required for accurate interpolation of the migration velocity field (Fig. \ref{fig:squareChannelFigure}), and inference of focusing positions. Symmetry can be exploited to reduce the number of computations.  Channel simulations in panels A-B took 200-250 particle locations and about 1 hour to be solved.  However, simulations in panel D took only $\sim$15 minutes on a laptop computer when taking advantage of 2 reflection symmetries, so that only one quarter of migration velocities needed to be computed, while panel C took $\sim$30 minutes when taking advantage of 1 dimension of symmetry, so that only one half of migration velocities were computed.}
     \label{fig:showOff}
 \end{figure}

 We have implemented the algorithm described in this paper in Matlab (Mathworks, Natick, MA),  and the resulting package (called \href{https://samuelechristensen.github.io/InFocus/}{\emph{INFOCUS}}) is freely downloadable from the first author's website. \textit{INFOCUS} is intended to be usable by builders of inertial microfluidic devices, can be run on a laptop computer, and needs minimal programming experience. It is designed for building quick predictions of where and how fast particles will focus inside straight microfluidic channels with arbitrary cross-section shapes. Users need only need define channel shape by specifying vertices for polygonal-shaped channels or a level-set function for curved boundaries, and specify their channel Reynolds number based on the maximum flow velocity, or on the flow rate. \textit{INFOCUS} can then calculate particle focusing positions, particle trajectories, and basins of attraction.  The tool is capable of calculating the focusing positions of channels with simple geometries on a laptop computer with 8GB of RAM in under an hour, while the most complex channel geometries that have assayed can be run on a desktop computer with 64GB of RAM in 1-2 hours (Fig. \ref{fig:showOff}). The implementation of our algorithm is able to handle arbitrary shapes and its speed makes parameter sweeps or shape optimization feasible.
 
 \section*{Acknowledgement}
 The authors acknowledge financial support from the National Science Foundation through grants DMS-1351860 and DMS-2009317. We are grateful to Prof. Claudia Falcon, Izak Oltman, Siting Liu, Dominic Diaz, Alejandro Diaz, and Prof. Hangjie Ji, for scientific discussions.

 \appendix
 \section{Discontinuity Solution} \label{sec:discontinuitySolve}
 We wish to solve Eqn \ref{eq:u11inner} analytically by transforming the problem into 3 separate inhomogenous Laplace equations and solving those using Spherical Harmonics, denoted $Y_i^j$, which have the property that 
 \begin{align}
     \Delta \left(r^nY_j^i(\theta,\phi)\right) = \left(n(n-1)-j(j+1)\right) r^{n-2}Y_j^i (\theta,\phi)
 \end{align}
 making them pseudo-eigenvectors of the Laplacian \cite{courant2008methods}. All code is in Mathematica (Wolfram Research, Urbana Champaign, IL)  and is based on code from \cite{hoodGitRepo}.
 We make the Helmholtz decomposition:
 \begin{align}
     (\boldsymbol\gamma \mathbf{\cdot \mathbf{x}\mathbf{e}_3}) \cdot \nabla \mathbf{u}_1^{(0)} +  \mathbf{u}_1^{(0)}\cdot \nabla  (\boldsymbol\gamma \mathbf{\cdot \mathbf{x}\mathbf{e}_3}) = \nabla \Phi + \nabla\times\mathbf{A} 
 \end{align}
 Our goal is not to calculate $\mathbf{u}_1^{(1)}$ completely, but only to isolate the terms that are $O(r^0)$ as $r\to\infty$. Throughout our calculation, we will isolate only terms that contribute at this order.
 
 We take the divergence of both sides of Eqn. \ref{eq:u11inner}, noting that since $\mathbf{u}_1^{(1)}$ is divergence-free:
 \begin{eqnarray}
     \Delta p_1^{(1)} & = & \Delta \Phi  = \nabla\cdot \left( (\boldsymbol\gamma \mathbf{\cdot \mathbf{x}\mathbf{e}_3}) \cdot \nabla \mathbf{u}_1^{(0)} +  \mathbf{u}_1^{(0)}\cdot \nabla  (\boldsymbol\gamma \mathbf{\cdot \mathbf{x}\mathbf{e}_3}) \right) \nonumber \\ &=& \nabla\cdot \left( (\boldsymbol\gamma \mathbf{\cdot \mathbf{x}\mathbf{e}_3}) \cdot \nabla \mathbf{u}_{\rm str} +  \mathbf{u}_{\rm str}\cdot \nabla  (\boldsymbol\gamma \mathbf{\cdot \mathbf{x}\mathbf{e}_3}) \right) + o(r^{-3})~.
     \label{eq:pPhiSolve}
 \end{eqnarray}
 Using the fact that spherical harmonics make a perpendicular basis, we rewrite the first term of Eqn. \ref{eq:pPhiSolve} as a sum of spherical harmonics. $\sum_{j=0}^\infty \sum_{i=-j}^{j} a_{ij}r^{-3} Y_j^i$, where:
 \begin{equation}
     a_{ij} = \int_0^\pi \int_0^{2\pi}\nabla\cdot \left( (\boldsymbol\gamma \mathbf{\cdot \mathbf{x}\mathbf{e}_3}) \cdot \nabla \mathbf{u}_{\rm str} +  \mathbf{u}_{\rm str}\cdot \nabla  (\boldsymbol\gamma \mathbf{\cdot \mathbf{x}\mathbf{e}_3}) \right) Y_j^i \sin \theta\, d\phi\, d\theta\nonumber
 \end{equation}
 We find a particular integral for this equation using the fact that spherical harmonics diagonalize the Laplace operator and lead to simple inversion 
 \begin{equation}
    p_\Phi = \sum_{j=0}^\infty \sum_{i=-j}^j r^{-1} n^\Phi_{ij} Y_i^j ~~\hbox{where}~~ n^\Phi_{ij} = \frac{ a_{ij}}{2-j(j+1)}~.
 \end{equation}
 Our first step toward calculating a leading order expression for $\mathbf{u}^{(1)}_1$  is to find a velocity field, $\mathbf{u}_\Phi$ associated with $p_\Phi$, and satisfying the equation:
 \begin{equation}
     \Delta \mathbf{u}^\Phi = (\boldsymbol\gamma \mathbf{\cdot \mathbf{x}\mathbf{e}_3}) \cdot \nabla \mathbf{u}_1^{(0)} +  \mathbf{u}_1^{(0)}\cdot \nabla  (\boldsymbol\gamma \mathbf{\cdot \mathbf{x}\mathbf{e}_3}) + \nabla p_\Phi \label{eq:uphi}
 \end{equation}
 We decompose each Cartesian component of the right hand side of \eqref{eq:uphi} into spherical harmonics: $\sum_{k=1}^3 \sum_{j=1}^\infty \sum_{i = -j}^j b_{ijk} Y_j^i \mathbf{e}_k$ where:
          \begin{align} b_{ijk} = \int_0^\pi \int_0^{2\pi}\left( (\boldsymbol\gamma \mathbf{\cdot \mathbf{x}\mathbf{e}_3}) \cdot \nabla \mathbf{u}_1^{(0)} +  \mathbf{u}_1^{(0)}\cdot \nabla  (\boldsymbol\gamma \mathbf{\cdot \mathbf{x}\mathbf{e}_3}) + \nabla p_\Phi \right)\cdot \mathbf{e}_k Y_j^i \sin\theta\,d\phi \,d\theta
     \end{align}
 We can then invert the Laplacian algebraically:
$\mathbf{u}_\Phi = \sum_{k=1}^3 \sum_{j=0}^\infty \sum_{i=-j}^j  \alpha^\Phi_{ijk} Y_i^j\mathbf{e}_k$ where   $\alpha^\Phi_{ijk} = -\frac{ b_{ijk}}{j(j+1)}$.
  
$\mathbf{u}_\Phi$ is not the solution at O(1) in the inner region because it has non zero divergence and doesn't satisfy the boundary conditions, this makes sense as $p_\Phi$ is not the full pressure.  To remedy this we simply add a potential to the pressure that would remove this divergence $p = p_\Phi + \nabla\cdot \mathbf{u}_\Phi$ and get a new RHS for a Laplace equation to represent with spherical harmonics
 \begin{align}
    \Delta \mathbf{u}^{(1)}_1 - \nabla (p_\Phi+\nabla \cdot \mathbf{u}_\phi) &= \left( (\boldsymbol\gamma \mathbf{\cdot \mathbf{x}\mathbf{e}_3}) \cdot \nabla \mathbf{u}_1^{(0)} +  \mathbf{u}_1^{(0)}\cdot \nabla  (\boldsymbol\gamma \mathbf{\cdot \mathbf{x}\mathbf{e}_3}) \right)\\
    \sum \sum \sum c_{ijk}\mathbf{e}_k &= \left( (\boldsymbol\gamma \mathbf{\cdot \mathbf{x}\mathbf{e}_3}) \cdot \nabla \mathbf{u}_1^{(0)} +  \mathbf{u}_1^{(0)}\cdot \nabla  (\boldsymbol\gamma \mathbf{\cdot \mathbf{x}\mathbf{e}_3}) + \nabla p_\Phi \right) \nonumber \\
          \text{where  } c_{ijk} =  \int_0^\pi \int_0^{2\pi} &\left( (\boldsymbol\gamma \mathbf{\cdot \mathbf{x}\mathbf{e}_3}) \cdot \nabla \mathbf{u}_1^{(0)} +  \mathbf{u}_1^{(0)}\cdot \nabla  (\boldsymbol\gamma \mathbf{\cdot \mathbf{x}\mathbf{e}_3}) + \nabla \left(p_\Phi + \nabla \cdot \mathbf{u}_\phi \right)\right)\cdot \mathbf{e}_k Y_i^j d\phi d\theta \nonumber
     \end{align}
          We can then perform one Laplace inverse for each of our 3 coordinate components:
     \begin{align}
     \implies \left(\mathbf{u}^{(1)}_1\right)_k =  \sum \sum \alpha_{ijk} Y_i^j\mathbf{e}_k \\
    \text{where      }   \alpha_{ijk} = -\frac{ c_{ijk}}{j(j+1)} \nonumber
 \end{align}
 Converting back to Cartesian coordinates we have our answer
 \begingroup\makeatletter\def\f@size{7}\check@mathfonts
 \begin{align} \label{eq:discontinuitySolution}
(\mathbf{u}^{(1)}_1)_x &= \frac{-5 \text{$\gamma $}_x^2 x \left(4 x^4+x^2 \left(7 y^2+z^2\right)+3 y^2 \left(y^2+z^2\right)\right)- 10 \text{$\gamma $}_x \text{$\gamma
   $}_y y \left(3 x^4+5 x^2 y^2+2 y^2 \left(y^2+z^2\right)\right) \ldots}{72 \left(x^2+y^2+z^2\right)^{5/2}} \nonumber \\
   &\ldots \frac{+5 \text{$\gamma $}_y^2 x \left(x^4+x^2 \left(y^2+z^2\right)+3 y^2
   z^2\right)}{72 \left(x^2+y^2+z^2\right)^{5/2}} \nonumber\\
  (\mathbf{u}^{(1)}_1)_y &= \frac{5 \text{$\gamma_x $}^2 y \left(x^2 \left(y^2+3 z^2\right)+y^2 \left(y^2+z^2\right)\right)-10 \text{$\gamma_x$} \text{$\gamma_y$} x
   \left(2 x^4+x^2 \left(5 y^2+2 z^2\right)+3 y^4\right)\ldots}{72 \left(x^2+y^2+z^2\right)^{5/2}} \\
  & \ldots \frac{-5 \text{$\gamma_y $}^2 y \left(3 x^4+z^2 \left(3 x^2+y^2\right)+7 x^2 y^2+4
   y^4\right)}{72 \left(x^2+y^2+z^2\right)^{5/2}}\nonumber \\
   (\mathbf{u}^{(1)}_1)_z &= \frac{5 z \left(-6 \text{$\gamma_x$} \text{$\gamma_y$} x y \left(x^2+y^2\right)+\text{$\gamma_x $}^2 \left(5 z^2 \left(x^2+y^2\right)+3 y^2
   \left(x^2+y^2\right)+2 z^4\right)+\text{$\gamma_y $}^2 \left(5 z^2 \left(x^2+y^2\right)+3 x^2 \left(x^2+y^2\right)+2 z^4\right)\right)}{72
   \left(x^2+y^2+z^2\right)^{5/2}}\nonumber
 \end{align}
 \endgroup
 




\section{Regularizing the Stresslet} \label{sec:regStresslet}
Formally the stresslet solves a $Re=0$ form of the Navier-Stokes equations:
\begin{equation} \mu \nabla^2 \mathbf{u}_{\text{\rm str}} -\nabla p_{\text{\rm str}} + \mathbf{f}_{\text{\rm str}} = \mathbf{0}~~,~~\nabla \cdot \mathbf{u}_{\rm str} = 0~, \label{eq:stressletPDE} \end{equation} for 
 \begin{equation} \textbf{f}_{\rm str} = -\frac{10 \pi \mu \gamma_x}{3} \left(\frac{\partial \delta(\textbf{x})}{\partial z},0,\frac{\partial \delta(\textbf{x})}{\partial x}\right)^{T} - \frac{10 \pi \mu  \gamma_y}{3}\left(0,\frac{\partial \delta(\textbf{x})}{\partial z},\frac{\partial \delta(\textbf{x})}{\partial y}\right)^{T}~, \end{equation} where $\delta(\mathbf{x})$ is the Dirac delta function and $(\gamma_x,\gamma_y)$ are the components of the shear vector. We follow \cite{cortez2001method} in deriving a non-singular approximation of $\mathbf{u}_{\text{\rm str}}$ in which $\delta(\mathbf{x})$ is replaced by a function, $\phi_{\epsilon}$;
 \begin{equation}
    \phi_{\epsilon}(r) = \frac{15 \epsilon^4}{8 \pi (r^2 + \epsilon^2)^{7/2}}
\end{equation}
for which an exact solution of Eqn \ref{eq:stressletPDE} is possible. For this choice of $\phi_\epsilon$, we may define auxiliary functions:
\begin{align}
    \nabla^2 G &= \phi_\epsilon & \nabla ^2 B &= G \label{eq:stressletauxiliary} \\
    H_1(r) &= r^{-1} B' - G & H_2(r) &= r^{-3} ( r B'' - B') \nonumber
\end{align}
where $(\cdot) = \frac{d(\cdot)}{dr}$, from which we can construct the components of the regularized stresslet:
\begin{equation}
    {u^{\epsilon}_{\rm str}}_i = \frac{H_2'}{r}x_i x_j x_k Z_{jk} + \left(H_2 + \frac{H_1'}{r}\right)x_k Z_{ik}~,
\end{equation}
where
 \begin{equation}
        \mathbf{Z} = \frac{10 a^3 \pi}{3}\begin{pmatrix}
        0 & 0 & \gamma_x \\
        0 & 0 & \gamma_y \\
        \gamma_x & \gamma_y & 0
        \end{pmatrix}.
    \end{equation}
We solve the system Eqn \ref{eq:stressletauxiliary} using Maple (Maplesoft, Waterloo, Canada), obtaining:
\begin{equation}      \textbf{u}^{\epsilon}_{\rm str} = \frac{-5a^3}{4  (\epsilon^2 + r^2 )^{5/2}}  \begin{pmatrix}
            2  xz(\gamma_x x + \gamma_y y ) + \epsilon^2 \gamma_x z \\
            2 yz(\gamma_x x + \gamma_y y ) + \epsilon^2 \gamma_y z \\
            2  z^2(\gamma_x x + \gamma_y y ) + \epsilon^2 (\gamma_x x + \gamma_y y) 
            \end{pmatrix}.
        \end{equation}

 \bibliographystyle{elsarticle-num}
\bibliography{references}



\end{document}